\documentclass[twoside]{elsart}

\usepackage{epsfig,graphicx}
\usepackage{subfigure}
\usepackage{amssymb}

\begin{document}
\begin{frontmatter}

\title{On the selection of AGN neutrino source candidates for
          a source stacking analysis with neutrino telescopes
        }

\author[Utrecht]{A.~Achterberg},
\author[Zeuthen]{M.~Ackermann},
\author[Christchurch]{J.~Adams},
\author[Mainz]{J.~Ahrens},
\author[PennPhys]{D.~W.~Atlee},
\author[Princeton]{J.~N.~Bahcall\thanksref{dec}},
\author[Bartol]{X.~Bai},
\author[BrusselsVrije]{B.~Baret},
\author[Dortmund]{M.~Bartelt},
\author[Irvine]{S.~W.~Barwick},
% ...
\corauth[cor]{{\scriptsize Corresponding author. Contact: gross@physik.uni-dortmund.de}}

% ...
\address[BrusselsVrije]{Vrije Universiteit Brussel, Dienst ELEM, B-1050 Brussels, Belgium}
% ...
\address[Christchurch]{Dept.~of Physics and Astronomy, University of Canterbury, Private Bag 4800, Christchurch, New Zealand}
% ...
\address[Dortmund]{Dept.~of Physics, Universit\"at Dortmund, D-44221 Dortmund, Germany}
\address[Irvine]{Dept.~of Physics and Astronomy, University of California, Irvine, CA 92697, USA}
% ...
\address[Mainz]{Institute of Physics, University of Mainz, Staudinger Weg 7, D-55099 Mainz, Germany}
% ...
\address[Bartol]{Bartol Research Institute, University of Delaware, Newark, DE 19716, USA}
\address[Princeton]{Institute for Advanced Study, Princeton, NJ 08540, USA}
% ...
\address[PennPhys]{Dept.~of Physics, Pennsylvania State University, University Park, PA 16802, USA}
% ...
\address[Utrecht]{Dept.~of Physics and Astronomy, Utrecht University/SRON, NL-3584 CC Utrecht, The Netherlands}
% ...
\address[Zeuthen]{DESY, D-15735, Zeuthen, Germany}
% ...
\thanks[dec]{Deceased}
\thanks[Berlin]{affiliated with Institut f\"ur Physik, Humboldt Universit\"at zu Berlin, D-12489 Berlin, Germany}
\thanks[Bari]{on leave of absence Universit\`a di Bari, Dipartimento di Fisica, I-70126, Bari, Italy}
\thanks[Kalmar]{affiliated with Dept.~of Chemistry and Biomedical Sciences, Kalmar University, S-39182 Kalmar, Sweden}

\begin{abstract}
The sensitivity of a search for sources of TeV neutrinos
can be improved by grouping potential sources together into generic classes in
a procedure that is known as  source stacking.  
In this paper, we define catalogs of Active Galactic Nuclei
(AGN) and use them to perform a source stacking analysis. The grouping of AGN
into classes is done in two steps: first, AGN classes are
defined, 
then, sources to be stacked are selected assuming that a potential neutrino
flux is   
linearly correlated with the photon luminosity in a certain
energy band (radio, IR, optical, keV, GeV, TeV). Lacking any secure detailed
knowledge on neutrino production in AGN, this correlation is
motivated by  hadronic AGN models, as briefly
reviewed in this paper.   

The source stacking search for neutrinos from generic AGN classes is
illustrated using the data collected by the AMANDA-II high energy neutrino 
detector during the year 2000. 
No significant excess for any of the suggested groups was found.  
\end{abstract}
\begin{keyword}
Neutrinos \sep AGN \sep source stacking \sep point sources\sep AMANDA \sep
IceCube 
\PACS 95.55.Vj\sep 95.85.Ry \sep98.54.-h\sep98.70.Sa 
\end{keyword}
\end{frontmatter}

\section{Introduction}
With AMANDA-II, different searches for high energy neutrino point sources have
been performed~\cite{tonioPS,paoloPS,4yrPS}, 
showing that neither the all sky search for hot spots in the neutrino
sky nor the search 
at predefined and astrophysically motivated source positions has led to
a discovery of a 
statistically significant neutrino signal. This however does not exclude the
possibility that 
the signal of a superposition of generically equal sources, each contributing
below the individual 
significance threshold, sums up to a significant signal for that specific
source type. In case 
of no detection, the stacked signal may be used to obtain a limit on neutrino
emission from that 
generic source type.
Source stacking methods have been applied in gamma-ray astronomy, e.g., in
searches for non-blazar AGN classes at GeV photon energies
\cite{cillis1,cillis2}. 

AGN belong to the most promising potential neutrino sources. Since the
detailed mechanism of neutrino production in AGN is still unknown, in this
paper a neutrino-production-model free 
attempt to classify the sources according to their geometrical properties and
electromagnetic 
emission is introduced.
Based on this classification, we develop an AGN stacking analysis to search
for TeV neutrinos.


In Sec.~\ref{classi}, we describe in detail the adopted systematic
classification of AGN based on an axisymmetric model \cite{urry}.
Possible neutrino production scenarios in AGN are reviewed in
Sec.~\ref{theory}. 
Pion production and subsequent decay results in correlated neutrino and photon
production. Neutrino fluxes
from these reactions are expected to be of the same order of magnitude as the
gamma-ray fluxes.
However, photons
interact in the source environment possibly leading to a
different photon spectrum than the neutrino spectrum. 
In Sec.~\ref{samples}, we define catalogs of AGN
classes based on our classification, and select from them 
interesting neutrino candidate sources. 
Hence, we order sources 
according to their photon flux at different energies.
In Sec.~\ref{opti}, a procedure is presented to determine the optimum number
of sources from each catalog to be included in the stacking analysis. 
This procedure is applied to determine source samples to be analyzed with the
AMANDA neutrino telescope, located at the South Pole~\cite{AMANDA}.
Thus selected sources in AMANDA's field of view are in the
Northern sky. Assuming 
hypothetical distributions of the source strength as a function of the ordered
source number 
we determine under which conditions 
a gain in sensitivity is obtained with the source stacking method
with respect to other methods \cite{tonioPS,kirsten_icrc}. 
In Sec.~\ref{signal}, we illustrate our method by evaluating the signal from
the resulting source samples using a data set collected by AMANDA in the year
2000 \cite{tonioPS}.  

\section{AGN classification}
\label{classi}

\begin{figure}[hbt]
\centering{
\epsfig{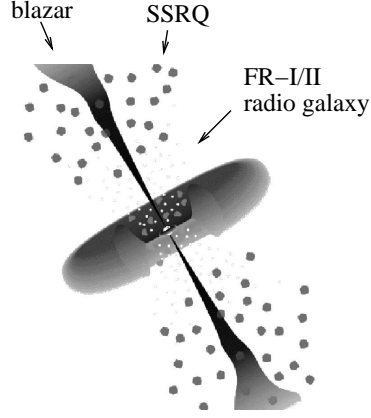}}
\caption{Scheme of an AGN with a black hole in the center and an accretion disk
perpendicular to the direction of two jets along its rotation axis.
The different inclination angles of the line of sight with respect to the jet
for blazars, steep spectrum radio quasars 
and radio galaxies are indicated by arrows. Image adapted from \cite{urry}. 
}
\label{AGNscheme}
\end{figure}

In this section we describe a possible classification of AGN useful to define
catalogs of sources to be stacked. 
Historically, a large variety of AGN (like Seyfert galaxies, radio galaxies or
quasars) have been  named due to
their appearance from Earth-based telescopes. 
The observational differences among the
various AGN types can be partially explained with a geometrically 
axisymmetric model as the  
result of different inclination angles, defined as the angle between the line of sight and the AGN axis
\cite{urry}. 
A pictorial scheme of an AGN showing the basic ingredients of the
axisymmetric model is shown in Fig.~\ref{AGNscheme}.
It consists of a rotating supermassive black hole, 
two jets with matter outflowing along the rotation axis and
an accretion disk of matter perpendicular to
the rotation axis. 
Aside from these geometrical differences, others will be outlined below. 

\begin{center}
\begin{figure}[hbt]
\epsfig{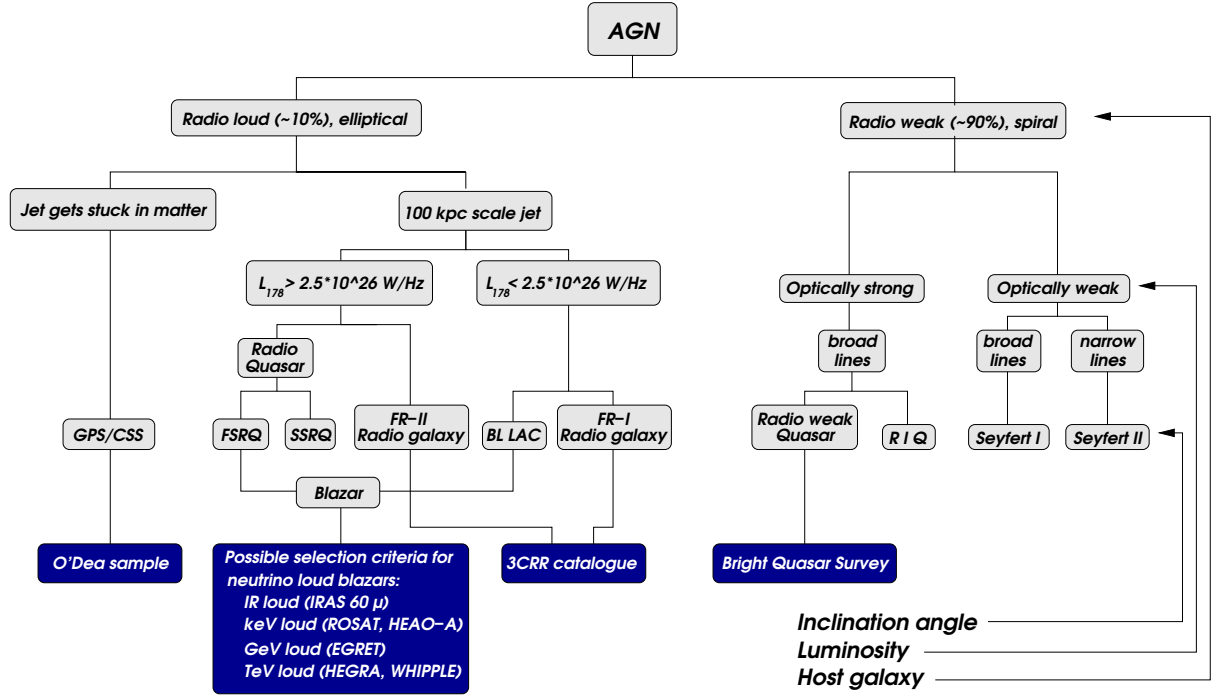}
\caption{AGN classification according to host galaxy, luminosity and 
inclination angle. }
\label{tree}
\end{figure}
\end{center}

We describe the AGN classification illustrated in Fig.~\ref{tree} based  on
the morphology of the host galaxy, the luminosity and the inclination
angle. AGN are generally divided into  radio-loud and radio-weak sources
as indicated in the first branching of the scheme.
Radio-loud AGN (left  branch of Fig.~\ref{tree}) can be classified according to
radio jet lengths: compact objects, where the jets get stuck in
dense matter \cite{odea}, are discussed with more detail in Sec.~\ref{gpscss}
and AGN with fully evolved jets of a length of  $100$ kpc up to
several Mpc in Sec.~\ref{L178}. 

Typically, AGN spectra are composed of a thermal part, 
the so-called blue bump with the maximum at optical-UV frequencies, and a
non-thermal part extending over up to 20 orders of magnitude in frequency. 
The blue bump is interpreted as thermal radiation of the warm inner 
accretion disk. The low energy component of the non-thermal spectrum 
in the radio to soft X-ray regime 
is assumed to be due to synchrotron radiation of electrons gyrating
in a magnetic field.
The origin of the high-energy component, if present, 
can be explained by hadronic or leptonic models (see Sec.~\ref{theory}). 
The non-thermal high energy photon emission is
known to be highly variable on a wide range of time scales, from less than one
hour to months \cite{Mrk421_orphan2}.
Within the thermal spectrum, some emission lines can be found.

Emission lines are classified as broad lines or as narrow lines.
The axisymmetric model assumes that broad lines are due to
fast moving dust clouds ($v \approx 1000$ km/s$-5000$ km/s) near the accretion
disk. On the other hand, narrow lines originate in slowly moving
dust clouds ($v \approx 500$ km/s) outside the torus. Depending on 
the angle of observation, the broad line emission from the clouds at the
center may or may not be hidden by the torus.
This model is confirmed by broad line observations in scattered
light from some AGN, which show in direct light only narrow lines
\cite{antonucci}. 

The radio emission of AGN is assumed to originate 
mostly in relativistic jets where it is caused by 
synchrotron radiation of electrons moving along the jet.
AGN are called radio-loud if the ratio of the radio flux at 5 GHz
to the optical flux is larger than 10 \cite{urry}.
Radio-loud AGN are located in elliptical galaxies, while radio-weak AGN are
located in spiral galaxies \cite{dowd}, and rarely in ellipticals.

Observations and jet models \cite{FBI} show a flat
radio spectrum for the flux density $F$ ($F \propto \nu^{\alpha}, \alpha > -0.5$)\footnote{The
spectral index $\alpha$  usually is determined between 2.7 GHz and 5 GHz.} for
the radio core,  
i.e.\ for the inner part of the jet. 
In contrast, radio lobes and hot spots 
located at the outer end of the jet typically 
show a steeper spectrum with spectral indices  from 
$-0.5$ to $-0.6$ in hot spots and from $-0.8$ to $-1.0$ in lobes.

For the radio-weak AGN (right branch of Fig.~\ref{tree}), a
luminosity dependent classification in optical wavelengths yields the 
division into quasars and Seyfert galaxies.
The intrinsically stronger objects are the radio-weak quasars
and the Radio Intermediate Quasars (RIQ) \cite{falckeRIQ}.
The radio-weak quasars are seen from moderate inclination angles 
($20^\circ-60^\circ$),
the RIQ are interpreted as the same objects, but seen from smaller angles. The 
radio emission of RIQ is relativistically beamed, similarly to blazars (see Sec~\ref{L178})
\cite{falckeRIQ}. 
The  weaker objects are the Seyfert galaxies,
classified as  Seyfert~I galaxies, if the core and the broad 
line region are visible,
or Seyfert~II galaxies, if the core is obscured by the torus \cite{unify_seyfert}.
Up to now, no Seyfert-like object with beamed emission has been observed.

\subsection{GPS and CSS}
\label{gpscss}

There is a substantial fraction of sources 
($\approx 1/3$ when selecting the strongest sources at 5 GHz) which shows, 
for frequencies above a certain turnover value, a steep radio spectrum from
the compact radio core.
If the turnover is in the MHz range, the source is classified as
\emph{Compact Steep Spectrum} source (CSS),
if it is in the GHz range, the source is a \emph{GHz Peaked Source} (GPS). 
These sources are significantly smaller than usual AGN: the
GPS show linear sizes below 1 kpc, the CSS in the range of 1-15 kpc.
GPS and CSS are not distinct source classes, since there is a 
continuous transition from the compact GPS to the slightly larger CSS.
As pointed out in \cite{odea}, the turnover frequency $\nu_m$ decreases with 
growing source size $l$ as
\begin{equation}
\nu_m \propto l^{-0.65}\,.
\end{equation}
The extremely high power of the GPS/CSS suggests a central engine similar
to those of other AGN to provide sufficient energy.

The source compactness can be explained by 
the assumption that jets get stopped by interactions with dense matter.
The most common interpretation of the turnover observed for
this source class is the model of  synchrotron self-absorption,
although free-free absorption on thermal electrons is not excluded. The
sharpness of the maxima found implies that the zone of the bulk radio 
emission is very small \cite{odeaBaum}, 
presumably located at the outer end of the jet.
GPS/CSS could be young states of
radio-loud AGN evolving into larger radio sources. The population 
statistics of GPS/CSS and extended radio-loud AGN are in agreement with 
this picture, provided they stay in the compact state for
a significant part of their lifetime.

For neutrino production, this class is of particular interest if
proton acceleration takes place within the central part of the jet.
As a matter of fact, the dense matter surrounding the source provides an ideal target for
pion production.
Since jets get stopped in the interaction zone, the predicted 
neutrino production is isotropic or only slightly beamed. Thus,
the inclination angle is considered to have only small effects on the
expected neutrino flux on Earth from this sources.

\subsection{Radio-loud AGN with 100 kpc scale jets}
\label{L178}

Two different jet morphologies have been observed for radio-loud AGN with
extended jets correlated with the radio luminosity at $178$ MHz~\cite{FR}.  
These AGN can be distinguished according to their luminosity:
the critical value of the luminosity is $L_{178}=2.5 \cdot 10^{26}$ W/Hz (see
Fig.~\ref{tree}),  
corresponding 
to a bolometric luminosity of $10^{46}$ erg/s \cite{FBII,FGB}. 
High luminosity AGN are characterized by powerful jets extending far
outside the host galaxy. The increasing jet luminosity at the outer end
produces extended radio lobes and the so-called hot spots.
These objects appear as Flat Spectrum Radio Quasars (FSRQ), as Steep Spectrum Radio
Quasars (SSRQ) or as FR-II radio galaxies~\cite{FR}.
Low luminosity AGN, divided in BL Lacs or as FR-I radio galaxies, have fainter jets. 
They show decreasing radio emission with 
growing distance from the central black hole and have no hot spots.

This classification of radio-loud AGN assumes  
a typical value of the Lorentz factor $\gamma \approx 10$ for
the bulk motion in the jet \cite{urry}.
This value can also explain the apparent superluminal motion of radio knots in
blazar jets \cite{mannheim}.

Radio-loud AGN show different appearances for different inclination angles. This is due to the
relativistic Doppler boost of the emission from the jet and the obscuration of
the inner core by the torus.
For large viewing angles to the jet axis (close to $90^\circ$), the torus obscures
the inner part of the AGN. Hence, the broad line region and 
the thermal continuum radiation of the accretion disk cannot be seen.
In this case, the AGN is called a \emph{radio galaxy}.  

If the opening angle of the torus is large enough, there is a range for the
inclination angle, where  the inner core is visible and the relativistic 
Doppler factor is smaller than $1$.
In this case the core is seen with its blue bump from the inner accretion
disk and broad emission lines are present in the spectrum. The radio spectrum of these objects is 
still steep and lobe-dominated.
Only high luminosity objects show this morphology, then  
appearing as SSRQ
with a bright optical core and strong broad emission lines.
The lack of similar low luminosity objects is still a matter of debate. 
Possibly, the inner core is obscured by the torus  
until the inclination angle is so small
that the emission becomes Doppler boosted \cite{FGB}.

For very small inclination angles 
($\lesssim 12^\circ$), 
the jet radiation is Doppler boosted 
due to the relativistic motion of the bulk outflow towards the observer.
These objects are characterized by a 
flat radio spectrum, strong variability and polarization. 
FSRQ and BL Lac objects can be combined into the blazar class since both are
characterized 
by strong beaming effects. 
The flat radio spectra of blazars can be 
explained by the dominance of the boosted flat-spectrum core over the 
non-boosted steep-spectrum radio lobes and will be used as selection criterion
for blazars. A flat spectral index in radio is an indication of 
optical thickness \cite{FBI}.

BL Lac objects are low-luminosity objects with a FR-I type jet,
while FSRQ are considered as the high luminosity objects (FR-II jet). 
The BL Lac objects are commonly divided into high-energy cutoff BL Lac 
(HBL) and low-energy cutoff BL Lac (LBL), referring to the maximum energy of
the electron synchrotron spectrum.
HBL are relatively weak in radio flux, strong in X-ray flux and bolometrically
less luminous than LBL. A distinction between these classes is usually made by
taking into account whether the radio-to-X-ray spectral index $\alpha_{rx}$ 
(with $F(\nu) \propto \nu^{\alpha_{rx}}$) is bigger (LBL)
or smaller (HBL) than -0.75 \cite{urry}.
The high energy $\gamma$ emission of HBL and LBL differs:
all confirmed AGN TeV sources are HBL\footnote{However, there are indications 
of TeV radiation from BL Lacertae (LBL) and M87 (misaligned BL Lac or FR-I)
 weakening the association.} \cite{tev}.
In contrast, the stronger AGN GeV (EGRET) sources 
consist of LBL and FSRQ, and only 2 of the 6 confirmed TeV $\gamma$ AGN 
have been detected by EGRET, with only moderate fluxes \cite{3EG}. 

\section{Selection of neutrino source candidates}
\label{theory}

Accelerated proton interactions with
ambient photons or matter lead to neutrino production
through the reactions: 
\begin{eqnarray*}
p \gamma &\rightarrow&  \Delta^{+}\rightarrow n \pi^{+}\\
p p &\rightarrow &  \pi^{\pm}+N\\
&&\pi^{\pm} \rightarrow \mu^\pm \nu_\mu \rightarrow e^\pm \nu_e \nu_\mu \nu_\mu
\,,
\end{eqnarray*}
and similar reactions for neutrons. 
These processes are always accompanied by neutral pion production. 
Neutrino and gamma ray fluxes are expected to be of the same order of
magnitude.
Since in most scenarios, proton and target photon
spectra fall steeply with energy, 
higher resonances and multi-pion production represent
a small correction to pion production.
The acceleration of the protons is assumed to be due to shock acceleration 
which may take place in the relativistic jet 
\cite{mannheim,biermannstrittmatter}
or in the accretion disk \cite{SteckerSalomon,nellen}.
Due to low plasma density in AGN jets, the $p\gamma$ interactions  
are likely to be dominant over $pp$ \cite{mannheim}.
This is also indicated by the lack of absorption lines
in AGN X-ray spectra \cite{SteckerSalomon2}. 
An exception from this consideration is given, if $pp$ interactions take 
place at the inner edge of the accretion disk \cite{nellen}.

If acceleration takes place in the disk, electromagnetic cascades  
initiated by photons from neutral pion decay would lead to
a non-thermal X-ray spectrum. 
However, measured AGN X-ray spectra and the 
diffuse X-ray flux show the dominance of a thermal flux peaked around 100 keV
\cite{LM}.  
This discrepancy might be solved by assuming that the non-thermal X-ray
spectrum contributes only $30\%$ to the measured X-ray flux 
\cite{SteckerSalomon2}. That implies that hadronic interactions are not
dominant. Hence, the resulting neutrino spectrum would be reduced by the same
factor. 
Due to this discrepancy, the disk model is disfavored as a dominant 
process for neutrino production.

For the jet as well as for the disk, 
the theory of diffusive shock acceleration suggests a power law spectrum with a
differential spectral index 
$\alpha \approx 2$ for protons \cite{bell1,bell2,prtheroe_SI}. Neutrinos
and photons are produced with the same index if multi-pion production is not
dominant.  
However, 
the energy of photons produced by neutral pion decay
may be redistributed to photons of lower energy by synchrotron pair
cascades. The photon escape energy depends on the  
optical depth. 
Hence, the spectra of neutrinos and photons can differ considerably, 
while the bolometric luminosity
remains correlated \cite{MPR}.

Hadronic jet models have been proposed to explain the
high energy $\gamma$ emission from the jet. 
A comparison to purely leptonic models can be found in
\cite{mannheim}.
Purely leptonic models usually explain
the high energetic (TeV) radiation by inverse Compton scattering of
photons. 
Models in which the photons originate as synchrotron radiation are classified
as Synchrotron Self Compton (SSC); those in which the photons are ambient are
classified as External Compton (EC).
The EC model requires high densities of external
photons to be Compton up-scattered to higher energies. These 
photon densities then constrain  the escape energy of  photons to values
below $300$ GeV.
Thus, the leptonic EC model cannot explain the observed TeV 
sources \cite{mannheim}. 
In the SSC models, the ratio between GeV and radio luminosity
is constrained, predicting a larger number of GeV blazars than observed by 
EGRET. Additionally, keV and TeV fluxes are correlated. 
Many observations confirm these
correlations and apparently favor the SSC model.
However, recent observations indicate the existence of orphan flares
for 1ES~1950+650 \cite{1ES_orphan} and for Mrk~421
\cite{Mrk421_orphan2,Mrk421_orphan},
where the TeV flux 
flares but the X-ray flux does not. 

In hadronic models, the emission of high energetic photons in the jet
 is not assumed to be directly from photohadronic processes. 
A non-negligible optical depth makes TeV photons cascade to lower energies.
Thus, an additional mechanism for high energy gamma
production is necessary. 
In proton initiated cascade models (PIC), it is assumed that electrons
scatter on some target photons acquiring higher energies.
In the SS-PIC (self synchrotron PIC) model, the target photons are those
emitted in the cascades.
The EC-PIC (external Compton PIC) assumes external photons to be scattered up 
to higher energies by the inverse-Compton process \cite{mannheim}.

Another possibility to explain observed high energy $\gamma$ emissions 
is given 
by the  Synchrotron-Proton Blazar model proposed by M\"ucke and Protheroe
\cite{MPro2}. It
assumes relativistic protons to emit synchrotron radiation 
at higher energies than the one from gyrating electrons.  
This model explains the observed double 
hump energy spectrum and predicts neutrino emission spectra 
for HBL and LBL \cite{MPro}.
The emission of TeV $\gamma$-rays via synchrotron radiation requires
protons to  
reach extremely high energies, which is possible only if the interaction rate 
is sufficiently low.
Hence, in TeV emitters $p\gamma$ and $pp$
interactions must be less frequent than in blazars without intense TeV photon
emission. This results in predictions of neutrino fluxes for HBL being 6
orders of magnitude less than for LBL.

In summary, to select source candidates, 
we assumed the common origin of neutrinos 
and photons from pion decay and the classification described before.
Since the optical depths of AGN 
are unknown, escaping photons can be of significantly lower energy.
This is taken into account by the selection of source candidates at
various energies where data are available
(radio, IR, optical, keV, GeV, TeV). 
Moreover, we assume that
the observed photon flux is proportional to the
TeV neutrino signal. 
While this selection does not depend on detailed AGN models,
other explicitely model dependent selections are possible, e.g.\
based on the concept of the jet-disk symbiosis \cite{FBI,FBII} used in
\cite{julia_paper}.

\section{Selection of neutrino candidate sources from catalogs of AGN classes} \label{samples}

We consider catalogs of AGN classes and apply some selection rules,
described in Tab.~\ref{table_source_cuts}, in order to obtain 'statistically
complete catalogs' of sources in a well-defined part of the sky above 
a flux threshold. In this selection we did not consider the 
variability in time of emissions from AGN.
As a matter of fact, most of the available photon data were not collected
at the same time as the considered AMANDA data set. 
Future multi-wavelength campaigns and simultaneous collection of
photon and neutrino data will allow approaches that also account 
for time dependencies.

To all the considered catalogs we applied the
requirement that the galactic
latitude $b$ is larger than $10^\circ$ (except for radio-weak quasars,
where the catalog itself requires $|b|>30^\circ$). This cut
excludes the galactic plane in order to avoid biasing AGN 
samples with galactic sources. Moreover, a minimum declination of $10^{\circ}$
is required since we look for neutrino induced upgoing muons.
We also consider a threshold on the distance corrected flux, to exclude
intrinsically weak nearby sources which might
otherwise migrate into the catalog due to their distance. 
The  formula used for the correction is given in appendix \ref{lumidi}. 
We stress that some blazar sources have been removed from our catalogs 
by a cut on the distance corrected flux.
The sources that were removed only due this cut 
are listed in Table \ref{table_nearby}.
Among these there are some well-known nearby AGN, e.g. 
NGC 1275 would be in the IR-blazar and in 
both keV-blazar samples and Mrk 501 would be in both keV samples.
The occurrence of these sources in various samples confirms 
our assumption that these sources 
migrate into the samples due to their proximity and should not be
considered as generic sources. 
These nearby sources with such high fluxes 
would bias a stacking analysis, hence they should 
be analyzed as individual sources \cite{tonioPS}.

In the following sections we describe with more details the
catalogs for the considered classes of sources
in Tab.\ref{table_source_cuts} to which we apply selection rules. Then 
we consider for which of these selected sources there
exist measurements of the photon flux that we assume to be correlated to
neutrino fluxes and give our lists of candidate neutrino sources
for the various classes.

\begin{center}
\begin{table*}[!ht]
\centerline{
\begin{tabular}{c|c|c|c}
Sample&  Flux /luminosity cuts & Coordinate cuts& Further cuts\\
\hline
Blazars& $F_{5GHz}> 0.8$Jy,  $F_{5GHz}^{z=0.1}> 1.5$ Jy& $\delta> 10^\circ, 
|b| > 10^\circ$&$\alpha > -0.5$ with $F_\nu \propto \nu^{\alpha}$\\
CSS & $ F_{178MHz} >10$Jy, log$P_{178MHz}> 26.5$ & $\delta>
10^\circ, |b| > 10^\circ$&Lin.\ size$ < 20 $kpc\\
GPS& $F_{5GHz}> 1$Jy, log$P_{5GHz}> 26.5$&$ \delta> 10^\circ,
|b| > 10^\circ$&$0.4 $GHz$< F_{max}< 6 $GHz\\
FR-I&  $F_{178Mhz} >10$Jy, & $\delta> 10^\circ,
|b| > 10^\circ$&\\
&$L_{178MHz}<2.5 \cdot10^{25}$ W/Hz&&\\
FR-II&  $F_{178Mhz} >10$Jy, & $\delta> 10^\circ,
|b| > 10^\circ$&\\
&$L_{178MHz}>2.5 \cdot10^{25}$ W/Hz&&\\
QSO& $B<16.16$, $U-B< -0.44$&$\delta> 10^\circ,
|b| > 30^\circ$&\\
\end{tabular}
}
\vspace{0.5cm}
\caption{
\label{table_source_cuts}
The selection criteria defining catalogs of the different source classes. $F_\nu$
stands for the photon flux at a certain frequency $\nu$ in Jy, $F_\nu^{z=0.1}$
is the distance corrected flux to a redshift of $z=0.1$, log$P$ is the
logarithmic power in W/Hz and $L$ is the luminosity. The declination is
labeled as 
$\delta$ while $b$ stands for the galactic latitude. For the blazars, the
spectral index $\alpha$ is determined between $2.7$ GHz and $5$ GHz. 
For CSS selected at $2.7$ GHz, a back-extrapolation of a steep spectrum
component to $178$ MHz was used instead of the measured $178$ MHz flux.
$U$ is the magnitude of the UV flux 
($\lambda=300$ nm$-400$ nm), $B$ the magnitude of the  flux at $\lambda=400$
nm$-550$ nm (blue). 
}
\end{table*}
\end{center}

\subsection{Blazars}

As explained in Section~3, blazars are characterized by a flat radio
spectrum, which we use as the selection criterion here to create
a 'blazar list'.
For radio sources with high fluxes at 5 GHz ($\lambda = 6$ cm),
K\"uhr et al.\ \cite{kuehr1Jy} worked out 
a complete catalog of radio sources with flux larger than about
1 Jy\footnote{$1 \rm{Jy} = 10^{-26} \rm{Ws}^{-1} \rm{m}^{-2}$, 
 in older papers also called 
f.u.\ for flux unit.}. 
A second version also covers  sources below this threshold, using data 
available up to 1981 \cite{kuehr2}. 
In the Northern sky, this catalog contains 
the ``strong surveys'' S1-S5 with a threshold of $0.8$ Jy and below. 
The selection rules select from the blazar catalog about 150
sources. 

\begin{figure}[!h]
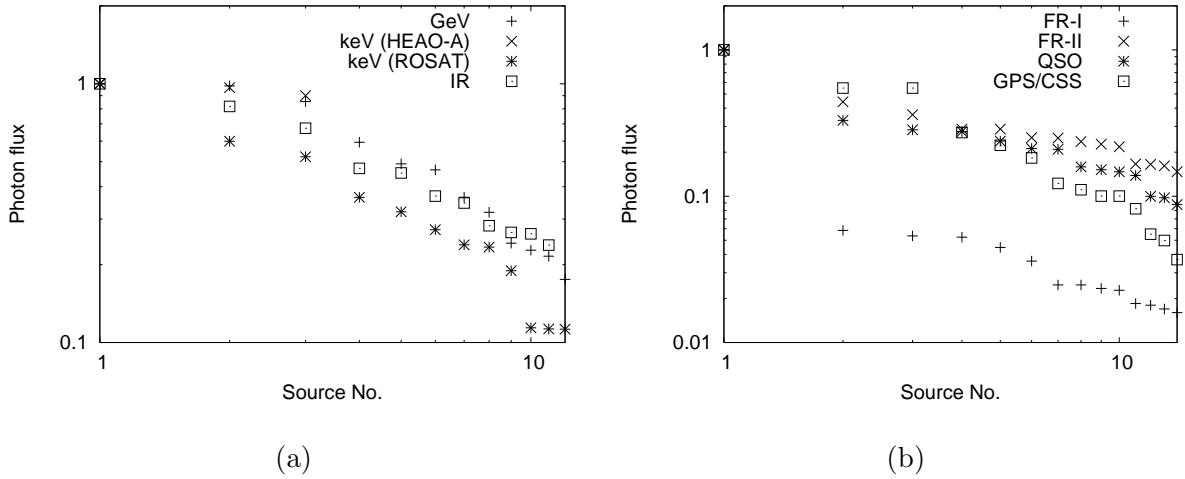

\centerline{
\subfigure[]{\epsfig{file=blazar_strength2.eps,width=8cm,angle=0}}
\subfigure[]{\epsfig{file=non_blazar_strength2.eps,width=8cm,angle=0}}}
\caption{Relative strength at the selection energy
of strongest sources in blazar samples (a) and in  non-blazar samples (b).}
\label{relstrength}
\end{figure}

Since blazar jets are in the direction of the observer, 
blazars are interesting candidate sources of neutrinos 
that could be produced in the jet.
Corresponding to different assumptions on the optical depth of blazar jets, the
neutrino flux is assumed to be proportional to the photon 
flux that are measured by various experiments in different energy ranges:
HEGRA and Whipple in the TeV one, EGRET in GeV, 
HEAO-A and ROSAT in keV or IRAS 
in infrared. 
The ordering of the sources is done according to the flux value
in the relevant energy range.
For the resulting samples, the relative source strengths at 
the selection energies are displayed in Fig.~\ref{relstrength}(a). 
The flux of the strongest source is normalized to 1.

For the sources in our list we look for
catalogs of the measured photon flux in various bands to
order them according to their photon emissions.

\paragraph*{IR-loud blazars:} the Infrared Astronomical Satellite (IRAS) 
provides a survey at wavelengths $12\mu$m,  $25\mu$m, $60\mu$m and $100\mu$m. 
The Faint Source catalog IRAS~F, covering most parts of the sky outside the 
galactic plane, includes 173,044 sources \cite{IRASf}. 
We find in total 12 sources in the IRAS F catalog that satisfy our selection
criteria and match sources in our original 'blazar list'.
The IR selected sample is given in Table~\ref{IR_table}.

\paragraph*{X-ray blazars:} in the X-ray regime, there are only a few 
all-sky surveys. We make the hypothesis that soft and hard X-ray emissions
are positively correlated with TeV neutrino production.
For hard X-rays (0.25-25 keV) there exist only the HEAO-A measurements
performed in the 1970s with
low sensitivity and relatively poor angular resolution.
This resulted in the 1H catalog \cite{1H} containing
842 sources.
The X-ray satellite ROSAT performed an all-sky survey in the 0.2-2 keV 
band, with high sensitivity and good angular resolution in the range of
a few tens of arc-seconds.
The strongest sources found in the all-sky survey are listed in the 1RXS
catalog \cite{1RXS}, which contains in total 18,811 sources.
For these antipodal satellites, two separate candidate lists are constructed
from 1RXS and 1H sources that match our 'blazar list'.
This resulting selections are listed in Table~\ref{heao-a_table} for HEAO-A
and in Table~\ref{rosat_table} for ROSAT.

\paragraph*{GeV blazars:} in the GeV range, the most sensitive data are those 
collected by the 
EGRET experiment on board the CGRO satellite~\cite{CGRO}. 
The third EGRET catalog (3EG) \cite{3EG}
contains in total 271 sources of which 66 have been identified as blazars and 
additional 27 sources with lower confidence. 
We select EGRET sources that are reliably
identified with a blazar in our list (class 'A' in the 3EG catalog), while 
weakly identified sources 
(class 'a' in the 3EG catalog) are accepted  
if an independent analysis  
confirms this identification \cite{EGRET_idSE}. The sources are sorted
according to the maximum flux detected by EGRET.
This yields the source subsample observable 
by AMANDA presented in Table~\ref{egret_table}.

We include a further sample (see in Table~\ref{3eg_unid_table}) selected
among the unidentified EGRET sources (these have
not yet been identified with sources detected at 
other wavelengths). 
Although a large fraction of these sources is likely to be galactic,
some may be extragalactic. To select presumably extragalactic
sources, aside from the usual galactic plane exclusion,   
we check that sources do not show features typical for galactic\footnote{Three
  different types of presumably galactic sources have been  
considered: two are concentrated on the galactic plane with 
$|b|<5^\circ$, while the 
third is associated with the Gould belt \cite{Gehrels}. The latter
form the so-called Local Gamma-Ray Population characterized by
constant fluxes below 
$2.4\cdot10^{-7}$s$^{-1}$cm$^{-2}$ and
a soft spectrum (photon spectral index $\gamma \approx 2.45$); they reach
galactic latitudes up to $30^\circ$.} sources.
Selected sources are listed in Table~\ref{3eg_unid_table}.
 These sources show a high variability, even higher than that of the
 identified AGN, and a large energy output that can be explained, even if they are
also AGN.  

\paragraph*{TeV blazars:}
up to now, only $6$ blazars have been firmly detected emitting TeV photons, of
which $5$ are in the field of view of AMANDA \cite{tev}. 
All TeV blazars show extreme variability in flux, such that a flux
ranking becomes impossible.
The sources all have a quite weak radio flux and do not
fulfill our radio selection criteria for blazars.
Furthermore, due to absorption by the IR background, only nearby sources can
contribute to the TeV photon flux.
These conditions prevent the use of our selection scheme for
TeV blazars. But since in optically thin sources, TeV photons are 
strictly correlated to TeV neutrinos, 
we consider all observable TeV blazars, listed in Table~\ref{TeV_table}, for
the source stacking analysis. 

\subsection{GPS and CSS}\label{CSSGPS}

For defining a catalog of compact sources, we follow the
selection of O'Dea \cite{odea}. The catalog contains former selections of CSS
by Fanti et al.\ \cite{fanti} 
and of GPS by Stanghellini et al.\ \cite{stanghellini}. 
Since CSS and GPS are characterized by a maximum in the radio flux at a
size-dependent frequency, the selection of these sources is done at several
frequencies: 178 MHz, 2.7 GHz and 5 GHz depending on the peak location. 

\begin{figure}[hbt]
\begin{center}
\epsfig{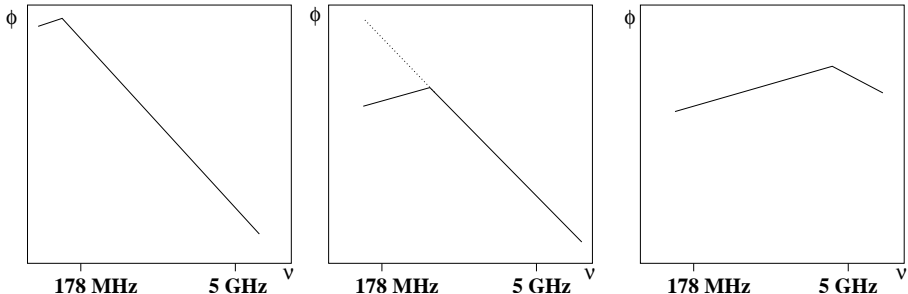}
\end{center}
\caption{Schematic plot of radio spectra of sources fulfilling the CSS and
GPS selection
rules.
On the left: a typical spectrum of a
CSS selected at $178$ MHz. In the middle:
a CSS spectrum with the maximum near $2.7$ GHz. On the right: a
GPS spectrum selected at 5 GHz. For the sources selected at $2.7$ GHz,
 the back-extrapolation of the steep spectrum component to $178$ MHz is
 relevant for the selection.
\label{schemeGPSCSS}}
\end{figure}

From the original CSS samples, all sources have been removed which 
do not fulfill all the selection criteria in the original paper 
\cite{fanti}.  The same has been done for CSS and GPS sources when
the requirements of the more recent analysis~\cite{odea} are not fulfilled.
Additionally, we put a cut on the absolute power at 5~GHz, 
similar to the  cut on the power at 178~MHz for the CSS sources indicated in 
Tab.~\ref{table_source_cuts}. The typical spectra of sources 
are plotted in Fig.~\ref{schemeGPSCSS}.
In total 40 sources are selected of which 9 are GPS, 
23 are CSS and 8 sources fulfill
selection rules for both classes.

Since CSS and GPS were selected at three different radio frequencies,
we cannot use the intensity in the radio. Since 
there are only few X-ray or IR data for these sources, the only 
possibility to get a suitable source ranking is to use data in the
optical band taken from \cite{odea}.
The underlying hypothesis we make is that the TeV neutrino flux 
is proportional to the optical flux.
Our resulting neutrino source catalog is
displayed in Table~\ref{table_CSSGPS}.

\subsection{FR-I and FR-II-radio galaxies}

As the distinction between FR type I and type II radio galaxies is made at
$178$ MHz, it is adequate to select radio galaxies 
from observations at this frequency.
Data at this frequency can be found in the 3C catalog in its revised version
(3CRR) \cite{3CRspinrad}.
The classification into FR-I and FR-II galaxies depends on the availability of
the redshift for the source and  is taken from \cite{3CRlaing}. 
The selection results in a catalog of $24$ FR-I and $122$ FR-II galaxies.
Additionally, $16$ compact sources fulfill the selection. 
They belong to the CSS source class as discussed in \ref{CSSGPS} above and 
are removed from the FR-I and FR-II samples.

For the selection of the neutrino candidate list, 
the radio flux at $178$ MHz is assumed to be proportional
to the TeV neutrino flux.
Hence, we use these fluxes to sort  the samples of radio galaxies.
The results are listed in Table~\ref{FRI} for
FR-I and in Table~\ref{FRII} for
FR-II galaxies.
Since FR-I/II radio galaxy jets do not point towards us,
the observable neutrino flux produced in the jets is expected to be weaker 
for these sources than for blazars. 
Nonetheless, neutrino production in the disk is expected to be 
isotropic compared to 
the case of jet emission. This enhances the observation probability
since it is not required that the observer is in the direction of the jet.

\subsection{Radio-weak Quasars}

A complete catalog of quasars selected at optical to UV wavelengths
is given by the \emph{Bright Quasar Survey} (BQS) \cite{SchmidtGreen}.
A quasar is identified by an excess of UV flux.
The catalogs presented by Sanders et al.\ \cite{Sanders} 
contain in total 109
sources, of which $59$ sources are radio-weak quasars in the sky viewable by
AMANDA.  

Multi-wavelength investigations of the BQS
sample show a second maximum in the photon spectrum 
at IR wavelengths for some of the quasars \cite{Sanders}. Hence,
in the usual assumption that photon flux features are correlated
to neutrino production at the source, we include in our list the
radio-weak quasars selected according to their photon flux at $60$ $\mu$m.
The full quasar catalog selected from the BQS according to the IR  flux 
is given in Table~\ref{table_bqs}.

For the resulting samples of CSS/GPS, FR-I/-II radio galaxies and 
radio-weak quasars, the relative source strengths at
the selection energies are displayed in Fig.~\ref{relstrength}(b).
The flux of the strongest source is normalized to 1.

\section{Optimization of the number of sources and the bin size}
\label{opti}

\subsection{The statistical optimization procedure}
\label{stat_meth}
The different hypotheses on the neutrino flux from the sources in our catalogs 
determine the relative source strength within each class but leave the
normalization free.  
We have normalized the signal by relating the sensitivity of the
point source analysis to the signal of the strongest source
in each sample. We consider here data taken by AMANDA during
the year 2000 \cite{tonioPS}.

In the point source analysis, a search was performed for a statistically
significant excess from any direction.
For this purpose, any position on a grid of
rectangular search bins is evaluated by
 considering the background expectation calculated
from the average number of events in the corresponding zenith range.  
No point sources were detected; the observed excesses were
equivalent to those in a randomized distribution.
In the considered data set, we find by multiplication of the
sensitivity\footnote{We define the sensitivity as the average upper limit in
  case of no  signal.} with the effective area, the lifetime of the analysis
\cite{tonioPS}  and the bin 
efficiency for a circular search bin of  $3.5^\circ$ radius that 
the sensitivity ($90\%$ C.L.) corresponds 
roughly to three neutrinos in a $3.5^\circ$ bin.
This motivates the choice of the normalization of the neutrino flux for the
most intense source that would produce a neutrino signal $S_1 \le 3$ events in
the considered AMANDA data set. 
Hence, for this source stacking analysis, we optimized the number of sources
using  $S_1=1, 2, 3$ events (also using the bin efficiency for 
$3.5^\circ$). 
Significantly higher  
signal normalizations would imply that the most intense source must be
  correlated with the position of a few spots, where an excess has been
  observed. These excesses are most probably
  due to statistic fluctuations. 
Since we do our selection independent of the results from that data set, 
  such a coincidence is not assumed. 
On the other hand, for $S_1<1$, the expected signal
  is too faint to result in a notable contribution in the source stacking
  analysis. 
For the small numbers considered, Poisson statistics has to be applied.
With our normalizations, the assumed neutrino fluxes for the sources in our
  catalogs agree with the limits of the current diffuse 
analyses \cite{kurt_nu2004,garyDA}. 

Since the sensitivity of the AMANDA-II detector is almost
independent of the declination, in the optimization procedure we approximate
the background  to be constant in the considered declination range. 
A total amount of 699 events in the year 2000 neutrino sample 
results in a background of $111.25/$sr for that year.

A given number of sources $N_{src}$ corresponds to a mean number
of expected signal events, $S$, and a mean number of expected background
events, $BG$.
Using Poisson statistics, we optimize for the median significance which will
be reached in $50\%$ of the 
experiments being performed under these assumptions.
First, we calculate the median number of events $n$ under 
signal hypothesis as a function of the number of sources to be included.
Then we calculate the Poisson
probability to observe at least $n$ events
with the assumption of pure background. Rescaled to the corresponding number
of standard deviations of a Gaussian, $\delta$, we get the median significance
as a function of the number of sources to be included. The optimum number of
sources, as determined by this median significance procedure, corresponds to
the highest value of $\delta$.  

\begin{figure}[htb]
\centerline{\epsfig{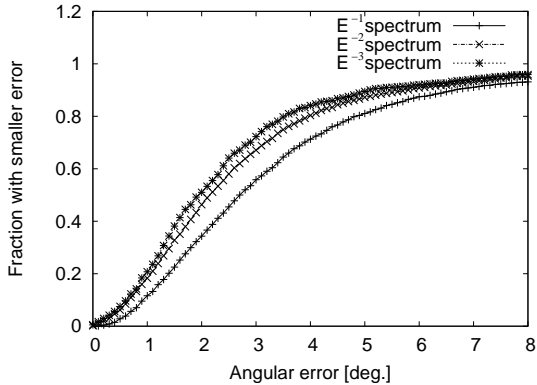}}
\caption{Cumulative Point Spread Function after quality cuts 
for various spectral indices in year 2000 sample.
}
\label{PSF}
\end{figure}

\begin{figure}[!h]
\centerline{\epsfig{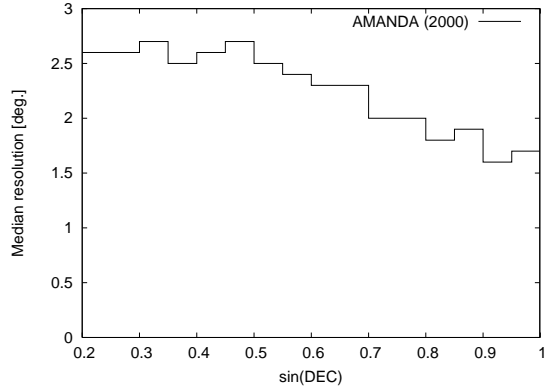}}
\caption[Median PSF as a function of the
zenith angle]{
Median PSF of AMANDA for data collected in 2000 as a function of
$\sin(DEC)$.
}
\label{PSF_vs_zenith}
\end{figure}

The same procedure is applied to the size of circular search bins, varying 
the signal according to the point spread function (PSF) and the background 
according to the area covered by the search bins.
The PSF is evaluated for the most likely case of an $E^{-2}$ spectrum. Since
the PSF is not completely independent of the declination, we use the PSF
averaged over all declinations. In
Fig.~\ref{PSF} the average PSF obtained by considering an isotropic source
distribution over the sky with $\delta>10^\circ$ 
is plotted for various spectral indices. The
declination dependence of the median of the PSF is shown  in
Fig.~\ref{PSF_vs_zenith} for an $E^{-2}$ spectrum. The resolution slightly
improved for larger values of the declination.

The optimum number of sources and the optimum bin size are not independent of 
each other. Thus, we follow an iterative 
procedure. First we evaluate the number of sources with rough estimates 
for signal and background. Then we optimize the
bin size for the given number of sources, and finally, we check the number of 
sources again.

The results for the optimum values have been checked by a second procedure
which will be called maximum observation probability procedure.
A minimum significance $\delta$ was predefined and the probability to observe
at least this significance is optimized. As the expected signal is rather
small, a value of $3 \sigma$ was chosen for $\delta$. 

\subsection{The resulting samples}
\label{results}
We use the total number of events and the PSF from the year 2000 point source
sample \cite{tonioPS} for the optimization to determine samples from our
catalogs to be analyzed for a cumulative neutrino flux. Our procedures may be
applied to other data sets easily. The resulting parameters for year 2000 
data are listed in Table~\ref{result}.

\begin{figure}
\centerline{
\epsfig{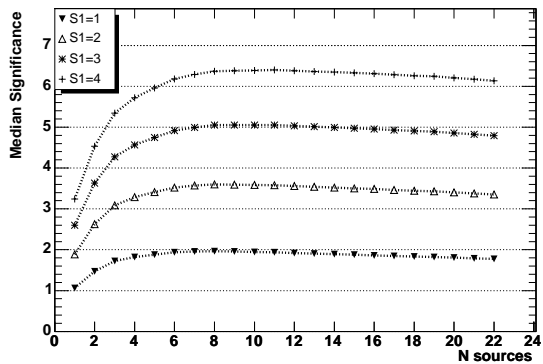}}
\caption{
Median significance as expected under the considered hypothesis as a function
of the number of GeV blazars to be stacked. The highest value 
of the median significance corresponds to the optimum number of sources.
\label{nsrc_egret}}
\end{figure}

In Fig.~\ref{nsrc_egret}, we plot the median significance as a function of the
number of EGRET blazars to be included 
for signal normalizations $S_1$ ranging from $1$ to $4$.
The highest normalization of $S_1=4$ signal events from the strongest source
is unlikely, as it implies a median
significance above $3\sigma$ for the single source search. Hence, $S_1=4$
 has not been taken into account when evaluating the optimum.

For most source classes, we find an optimum of about $10$ sources. 
With one exception,
 the source stacking analysis is more sensitive in testing the corresponding 
hypotheses than the point source analysis of single sources. 
Usually, the observed peaks are asymmetric: if more sources than the optimum
number are included, the expected significance falls only slowly. On contrast
the significance falls quite steeply if fewer sources are included.

The exceptions are found for FR-I and for FR-II radio galaxies, where the
results are completely different.
As there is a luminous FR-I galaxy in our neighborhood, M87, 
the whole FR-I flux would be strongly dominated by M87. 
The detection of TeV photons 
from this  
non-blazar AGN by HEGRA~\cite{HegraM87} and H.E.S.S.~\cite{HessM87},
theoretically predicted in \cite{DoneaM87},  also supports 
the uniqueness of this source.
Thus, the optimization 
suggests just to analyze the neutrino flux from that source only. In other
words, the analysis of the strongest source only
is more sensitive to our hypothesis than source stacking for the corresponding
catalog. 
However, for two reasons we decided to analyze additionally a FR-I sample
without M87. 
First, there is always the possibility that a single source 
does not contribute to the neutrino flux due to a reason not considered
at selection.
Second, in the opposite case, if M87 would be identified as a neutrino point
source, the cumulative signal of the other sources contains information
whether the M87 neutrino flux would be a specific feature of that source or a
general characteristic of FR-I galaxies. 
Thus, we analyze a FR-I sample without M87 additionally. 
Then we find an optimum value of $20$ sources to be included.

The situation of FR-II galaxies is opposite to that of FR-I galaxies. 
There are a lot of sources at higher redshifts with 
similar luminosities. This results in an monotonically increasing sensitivity
while including more sources. Eventually, the coverage of the sky by search
bins reaches a significant fraction of the sky
and the techniques of point source analysis become less suitable
to analyze the signal. If the energy spectrum of the signal differs
from  the atmospheric background, diffuse techniques searching for global
excess of high energy neutrinos should be more sensitive to FR-II galaxies.
However, since point source analysis and diffuse analyses are using
complementary techniques and a point source analysis can give more
information about the origin of the excess, a source stacking analysis of FR-II
galaxies may also be useful.
Since  our optimization fails, we have to define the most suitable value for
the number of FR-II sources to analyze.
For that, we use a local saddle point at
$17$ sources for the cutoff where the sensitivity grows only marginally
if further sources are included.

The unidentified EGRET blazars show  nearly constant values 
for observation probability and median significance between
$15$ and $35$ sources.

The hard X-ray sources are restricted to three sources due to the poor
sensitivity of HEAO-A, an experiment of the 1970s.

We cross-checked the procedure based on the mean significance to that based on
the maximum observation probability. The deviations between the results from
the two procedures are within 
the corresponding uncertainties of the determination of the optimum.

For the optimum bin size, we find results between $2.6^\circ$ for the
unidentified GeV sources and $3.0^\circ$ for the single source M87. As expected
\cite{alexandreas},  
the bin size decreases with the size of the sample.

\begin{table*}[!hbt]
\centerline{
\begin{tabular}{c|c|c|c|c}
\bf source class & \bf $N_{src}^{MS}$ 
&\bf $N_{src}^{DP}$ &  \bf listed in& \bf Bin size [deg.] \\
\hline
&&&\\
IR blazars (IRAS) &$11^{+0}_{-1} $ &$
11^{+0}_{-1} $& table \ref{IR_table}& $2.8\pm 0.2$ \\
keV blazars (ROSAT) &
$8 \pm 1$ &$9 \pm 1$& table \ref{rosat_table}& $2.8\pm 0.2$\\
keV blazars (HEAO-A)&$3 \pm 0 $ 
& $3 \pm 0 $ & table \ref{heao-a_table}&$2.9\pm 0.2$\\
GeV blazars & $8 \pm 1$& $9 \pm 1$ & table \ref{egret_table}& $2.8\pm 0.2$\\
unidentified GeV sources&$22\pm5$ & $20 \pm 5 $ & table 
\ref{3eg_unid_table}&$2.6\pm 0.2$\\
TeV blazars & $5^*$&$5^*$& table \ref{TeV_table}& $2.8\pm 0.2$\\
GPS and CSS & $8 \pm 1$ &$8 \pm 2 $& table \ref{table_CSSGPS}& $2.8\pm 0.2$ \\
\hline
FR-I radio galaxies&$1Ž \pm 0 $
&$1Ž \pm 0 $&table \ref{FRI}& $3.0\pm 0.2$\\
FR-I  without M87&$20^{+3}_{-5}$&$19 \pm 4$&table \ref{FRI}&$2.6\pm 0.2$\\
FR-II radio galaxies&$122$
&${17^{+105}_{-5}}^{**}$  &table \ref{FRII}&$2.6\pm 0.2^{***}$\\
Radio-weak quasars&$11^{+1}_{-3}$
&$11^{+1}_{-3}$&table \ref{table_bqs}& $2.8\pm 0.2$\\
\end{tabular}
}
\vspace{0.5cm}
\caption{Resulting parameters for the source stacking analysis.
\newline
$N_{src}^{MS}$ and  $N_{src}^{DP}$ stand for the optimum number of sources
determined with the median significance procedure and, respectively, the
maximum observation probability procedure.\newline
$ ^{*}$ All sources included without optimization, see above .\newline
$^{**}$ 
This value was taken for the analysis. Only a saddle point. The observation probability increases again,
when including all sources 
($\rightarrow$ diffuse analysis). See Sec.~\ref{results}.\newline
$^{***}$ Bin size evaluated for $17$ sources. 
\label{result}}
\end{table*}

\subsection{The relation between flux distribution and the  optimum 
number of sources}
As described above, the optimization of the number of sources can
lead to two special cases, where either the source stacking analysis
degenerates to a point source analysis of a single candidate source, or
to a diffuse analysis where the arrival direction of the neutrino becomes
irrelevant. 
If the flux falls only slowly from the
strongest source to the weakest,  
as is the case for FR-II radio galaxies, 
then we cannot find an optimum number of sources. 
So the probability to observe a signal always 
increases when more  sources are added. 
In the other extreme, as was found for FR-I  radio galaxies, 
if the strongest source is expected to 
contribute most of the flux and the other sources are substantially weaker,
then the sample of sources to be stacked degenerates to a single source. 
Here we investigate the conditions on the flux distribution
for these degeneracies to occur.
We consider the flux $F$  of the $N^{th}$ source, $F(N)$.
Motivated by the nearly linear decrease of $F(N)$ in the double 
logarithmic plots in Fig.~\ref{relstrength}, we assume a power law,
\begin{equation}
F(N) \propto N^\alpha\,.
\end{equation}
For $\alpha \ge -1$, the finiteness of the total
 flux requires a cut-off or a steepening.

For different values of $\alpha$, we evaluate the optimum number of sources
and  determine the range of values for $\alpha$, 
where the two degeneracies occur.
The maximum number of considered sources is set to $100$. This corresponds 
to a coverage of about $10\%$ of the sky viewable by AMANDA, assuming a bin
size of $2.6^\circ$ . For significantly 
more sources, the sky get densely covered by candidate sources and  the
directional information cannot be used  
 and diffuse methods are more useful.

For a slowly falling flux distribution with $\alpha> -0.65$, 
we cannot find an optimum number and all sources
have to be included. In the range $-0.65>\alpha>-2$, a non-degenerate
optimum number of sources 
is found and for $\alpha<-2$, only the strongest source
has to be selected. The thresholds depend on the detector parameters, although
for $\alpha>-0.5$, $S/\sqrt{BG}$ diverges, i.e.\
the degeneracy occurs generally for these values. 
An improved angular
resolution can explore the range  $-0.65<\alpha<-0.5$ for
a source stacking analysis.

A fit of $\alpha$ on the flux distributions of our samples results in
$\alpha=-0.6$ for FR-II radio galaxies and for IRAS blazars, 
a 2-point index $\alpha=-4.1$ for
the two strongest FR-I radio galaxies and $-0.7 \ge \alpha \ge -1.1$ 
for the other samples.
For all samples, the results of the optimization are in agreement with
our classification according to $\alpha$.

\section{The results for the year 2000}
\label{signal}
\begin{figure}[!ht]
\centerline{\epsfig{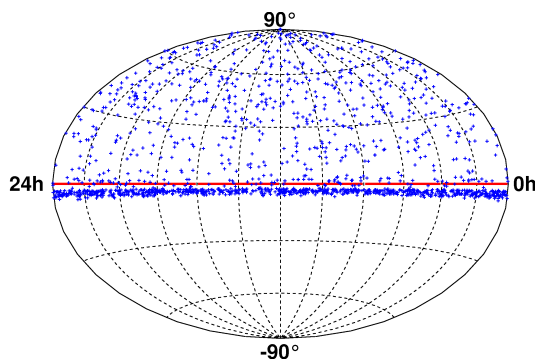}}
\caption[Reconstructed arrival direction of  neutrino events in
the year
2000-2003 sample]{Reconstructed arrival direction of $1557$ selected events in
  the year 2000 sample \cite{tonioPS}. The events below $\delta=0^\circ$ are
  dominated by   atmospheric muons.
}
\label{skyplot}
\end{figure}

\begin{figure}[!ht]
\centerline{\epsfig{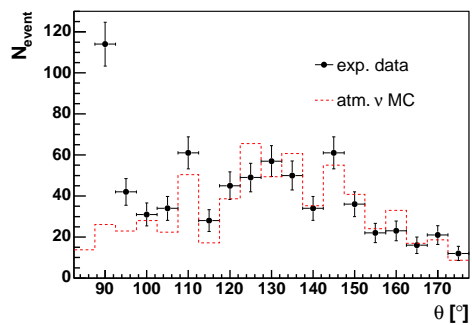}}
\caption
{Distribution of the zenith angle in the considered data set in comparison
  with the expectation of atmospheric neutrinos from simulations.
}
\label{zenith_distr}
\end{figure}

\begin{figure}[!ht]
\centerline{\epsfig{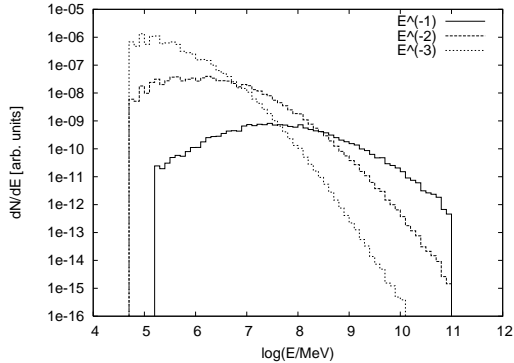}}
\caption
{Energy spectra folded with the response of the AMANDA detector for
  hypothetical source spectra with the spectral indices $-1$, $-2$ and $-3$.
  }
\label{dNdE_detector}
\end{figure}

The sources we selected  are listed in Table
\ref{egret_table}-\ref{table_bqs} in the appendix. 
We apply the source stacking method
to the data collected by AMANDA in the year 2000 using the event sample
from the previous point source analysis \cite{tonioPS} corresponding to a
detector lifetime of $197$ days. The selection 
of well reconstructed events was optimized for high sensitivity to point
sources. In Fig.~\ref{skyplot}, the reconstructed arrival
directions of the individual events of this data set are shown in a sky
map. The zenith distribution of these events is displayed in
Fig.~\ref{zenith_distr} in comparison with the simulated angular 
distribution of
atmospheric neutrinos. Since the event selection was not optimized for 
purity, the excess of measured 
events close to the horizon is expected as the contribution 
of poorly reconstructed atmospheric muons.
 The detector response with this event selection
was simulated for various spectral indices. The energy spectrum folded with
the Earth absorption factor and with the
detector response is shown in Fig.~\ref{dNdE_detector}. A uniform source
density on the Northern Sky was used in the simulation, corresponding to a
mean zenith angle of $33.4^\circ$.
For an $E^{-2}$ power
law spectrum, as expected from Fermi acceleration, that means $90\%$ of the
signal events are between the AMANDA energy threshold at $50$~GeV and
$350$~TeV. 

Results obtained on a more complete set of data collected in the years
2000-2003 will be reported in \cite{4yrPS}.
Moreover, this method and the list of sources we derived will serve as the
starting point for stacking analyses performed with the IceCube neutrino
telescope \cite{IceCube} which is currently under 
construction. AMANDA will be integrated into IceCube.
The source stacking method will be adapted to the better performance
of the larger detector in terms of effective area and angular
resolution. Application of this method for other neutrino telescopes
\cite{BAIKAL,ANTARES,NESTOR,NEMO,KM3} is conceivable.

In this analysis, only the cumulative result for a class of sources is
evaluated, not the signal for individual sources. 
The AMANDA collaboration follows a strict blindness policy, i.e.\
an analysis has to be developed blindly with respect to the data. This prevents
statistical fluctuations to affect the final steps of the analysis. 
In the context of the source stacking analysis, the selection of sources was
done independently of the AMANDA data itself. 
No signal was evaluated until
the source samples and all analysis parameters were fixed. 

The number of observed events in our search bins has to be compared with the
background, mainly atmospheric neutrinos. Since the background  is independent
of the right ascension, for each position in
the sky it can be evaluated from the events in the same declination band.
If two sources of the same sample are very close to each other, 
their search bins may overlap. Then, the background expectation is corrected
for the overlap and events in the overlapping area contribute only
once to the cumulative signal.

\begin{figure}[ht]
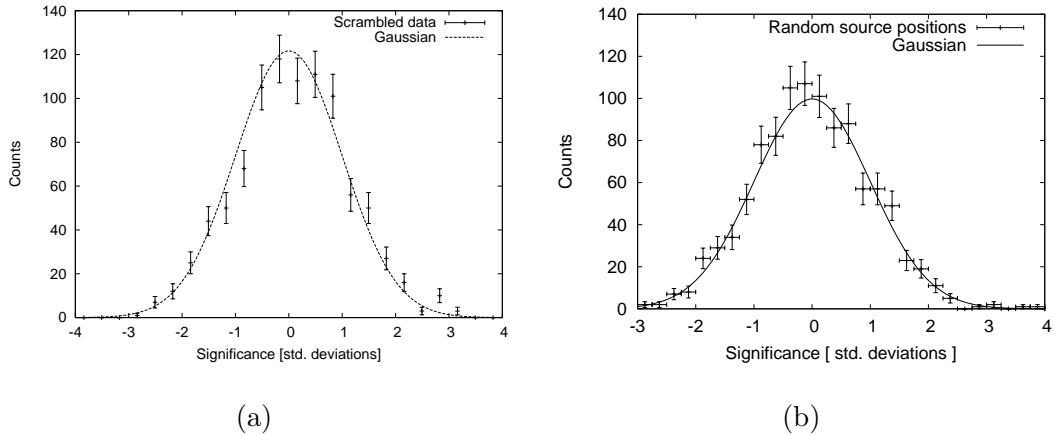

\begin{center}
\centerline{
\subfigure[]{
\epsfig{file=signi_scramble_2000.eps,width=7cm,angle=0}}
\subfigure[]{
\epsfig{file=significance_random_sources_bw_2000.eps,width=7cm,angle=0}}}
\caption{Significance distribution for selected sources in scrambled data sets
  (a) and for samples of 10 sources with random source positions evaluated
  with the unscrambled data set (b).}
\label{signi_random}
\end{center}
\end{figure}

The probability to observe the measured number of events under the hypothesis
of pure background is evaluated by Poisson statistics. 
The correctness of the analysis was
tested with two procedures. First, to test the correct evaluation of the
significance, a collection of data sets with randomized right 
ascension was created and the significance of the signal from the stacked
sources in these randomized data sets was determined. As expected for
randomized data sets, the significance 
follows a Gaussian with a width of one centered at zero.
Additionally,  hypothetical source 
lists with random source positions were evaluated using the original data
set. The results of the point source analysis of the considered data set
\cite{tonioPS} and the unfolded neutrino spectrum obtained from that data set
\cite{kirsten_icrc} suggest that the data sample is strongly dominated by
atmospheric neutrinos.
If this holds, a difference of the significance distribution to that of
randomized data sets, which follows a Gaussian (see above), would 
indicate that our assumption of a flat background in right ascension is
not correct. Also in this case, the observed significance distribution follows
our expectation.
The significance distribution for both tests is shown in
Fig.~\ref{signi_random}. 

\begin{figure}[!h]
\begin{center}
\epsfig{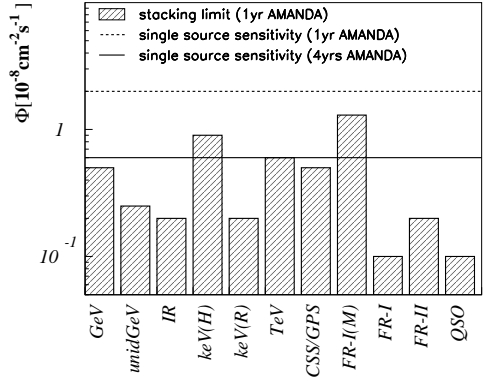}
\end{center}
\caption[Limits on the  neutrino flux from an average source using data from   the year 2000]{
\label{limit_plot_2000}
Limits on the neutrino flux from an average source in the generic AGN classes
  using data from   the year 2000. The average sensitivity to
  single point sources  is
  indicated by horizontal lines for this data set \cite{tonioPS} a well as for
 a 4 year data set \cite{4yrPS}.
}
\end{figure}

No statistically significant signal was found, in this limited data sample,
above pure background; only upper limits could be derived.
To calculate these limits we used Feldman-Cousins \cite{FC} confidence
intervals and Poisson statistics. 
Systematic errors, estimated to be at the  $\pm 30\%$ level, were not taken
into account.
The measured event rates, the background expectation and the resulting 
upper limits in terms of event counts and integral neutrino fluxes above
10 GeV are presented in 
Table \ref{table_limits} and visualized in Figure~\ref{limit_plot_2000}.  A dedicated analysis of systematics for a
multi-year dataset is currently in preparation.
For most of the declination range, the sensitivity ($90\%$ C.L.) of the point source
analysis for the considered data set is about  $2\cdot10^{-8} $cm$^{-2}$
s$^{-1}$ \cite{tonioPS}. A comparison with the limit per source obtained from
stacking (see
Table~\ref{table_limits}) shows the progress 
 reached by the source stacking method.  

\begin{center}
\begin{table}[!ht]
\centerline{
\small
\begin{tabular}{c|c|c|c|c|c|c}
sample & $N_{src}$ & $N_{\nu}^{obs}$ & $N_{\nu}^{bg}$ & $n_{lim}$ & $f_{lim}$
& $f_{lim}/N_{src}$\\
\hline
IR blazars & 11 & 7 & 10.17 & 3.0 & 2.0 & 0.18\\
keV blazars (ROSAT)  & 8 & 4 & 6.68 & 2.4 & 1.6& 0.2\\
keV blazars (HEAO-A) & 3 & 2 & 2.47 & 3.5 & 2.8& 0.9\\
GeV blazars & 8 & 6 & 5.3 & 6.3 & 4.0 & 0.5\\
unid.\ GeV sources& 22 & 15  & 14.9 & 7.6 & 5.6 & 0.25\\
TEV blazars & 5 & 4 & 4.53 & 4.1 & 2.8 & 0.56\\
GPS and CSS & 8 & 7 & 6.14 & 6.4 & 4.3 & 0.54\\
FR-I galaxies & 1 & 0 & 0.56 & 1.9 & 1.3 & 1.3\\
FR-I without M87 & 20 & 9 & 11.50 & 3.9 & 2.7 &0.14\\
FR-II galaxies& 17 & 10 & 13.42 & 3.7 & 2.7 & 0.16\\
radio-weak quasars& 11 & 4 & 7.55 & 1.9 & 1.3 & 0.12
\end{tabular}
}
\vspace{0.5cm}
\caption{
\label{table_limits}
Results for the year 2000 data: Number of sources $N_{src}$, measured number of
events $N_\nu^{obs}$,
the corresponding background $N_\nu^{bg}$ and the $90 \%$ C.L.\ limits on the
event counts ($n_{lim}$) and on the integral flux for an $E^{-2}$ spectrum
above 10 GeV ($f_{lim}$) in units of $10^{-8} $cm$^{-2}$
s$^{-1}$. $f_{lim}/N_{src}$ represents the limit per source.
}
\end{table}
\end{center}

\section{Summary}
We have performed a systematic classification of AGN. This is the basis of a
source stacking analysis of TeV neutrinos with the AMANDA
neutrino telescope. Neutrinos are assumed to be produced in optically thick
jets or accretion disks. The optical depths of these sites and consequently the
average energy of photons leaving the source are unknown.
We therefore select the sources according to the
photon flux at different energies. Most of the available photon data were not
collected at the same time as the considered AMANDA data set. Thus, the time
variability of AGN was not considered in this analysis.

With the hypothesis that the TeV
neutrino flux is  proportional to the corresponding photon flux,
we have optimized the number of sources to be included.

For most source classes,  an optimum of $8-12$ sources 
to be stacked has been found. 
For these, the source stacking analysis is more sensitive than
the point source analysis of a single source.
Exceptions have been found for FR-I and FR-II radio galaxies. 
While FR-I galaxies are dominated by the single source M87, there are
too many FR-II galaxies 
with a similar flux in order to converge at a reasonably small number of
sources to be stacked.

For the data of the year 2000, no significant deviation from the background
expectation was found. In a next step, the stacking method will be applied to
several years of AMANDA data. Optimizing the method to an angular resolution of
less than $1^\circ$ and better background rejection, it will be adapted to the
$1$ km$^3$ IceCube array \cite{IceCube}.
Starting with the selection of southern sky sources, it is also applicable to
other neutrino telescopes
\cite{BAIKAL,ANTARES,NESTOR,NEMO,KM3}.

\newpage  
\appendix
\section{The distance correction of fluxes}
\label{lumidi}
The distance dependence of a measured flux $F_\nu$ 
at the observer's frequency $\nu$ of a source with 
luminosity $L_{\nu^\prime}$ is given by 
\begin{equation}
F_\nu {\rm d}\nu=L_{\nu^\prime} \frac{1}{4 \pi d_l(z)^2}{\rm d}\nu^\prime\,,
\end{equation}
with the luminosity distance $d_l$ and 
the frequency at the source $\nu^\prime=\nu(1+z)$.
The luminosity distance as a function of redshift is given by
\begin{equation}
d_l (z) = \left \{ 
\begin{array}{lc}
\frac{(1+z)c} { H_0 \cdot | \Omega_k |^{0.5} } 
\, {\rm sin} \left\{ {\left| \Omega_k \right|}^{0.5} \cdot I(z) \right\}
\,, &\mbox{ if } \Omega_k < 0\,,\\
\frac{(1+z)c} { H_0  } \,,
&\mbox{ if } \Omega_k=0\,,\\
\frac{(1+z)c} { H_0 \cdot | \Omega_k |^{0.5} } 
\, {\rm sinh} \left\{ {\left| \Omega_k \right|}^{0.5} \cdot I(z) \right\}
\,, &\mbox{ if } \Omega_k > 0\,,
\end{array}
\right.
\end{equation}
see \cite{ld}.
$I(z)$ is given by
\begin{equation}
I(z) = \int_0^z \left [ (1+z^\prime )^2 \cdot (1 + \Omega_m z^\prime) - 
\Omega_\Lambda z^\prime (2+z^\prime) \right ] ^{-0.5} dz^\prime\,,
\end{equation}
with the normalized curvature $\Omega_k$.
We used  $\Omega_m=0.27$, $\Omega_\Lambda=0.75$ according to \cite{Spergel}
and  $h  = 0.71$ according to \cite{Balbi}.

For a power law spectrum with spectral index $\alpha$, the redshift results in
\begin{equation}
L_{v^\prime} {\rm d}{\nu^\prime}= L_{\nu\cdot(1+z)} \cdot (z +1)^{\alpha+1}{\rm d}\nu\,.
\end{equation}
Finally, the correction to $z=0.1$ for a source at redshift $z$
  is given by
\begin{equation}
F^{z=0.1} = F\cdot \left ( \frac{d_l(z)}{d_l(z=0.1)}\right )^2\cdot
\left( \frac{z+1}{0.1+1} \right)^{\alpha+1} \,.
\end{equation}

\section{Source catalogs for AMANDA}
\begin{center}
\begin{table}[htb]
\begin{footnotesize}
\centerline{
\begin{tabular}{c|c|c|c|c|c|c|c|c|c}
\bf radio source & \bf IR source &  \bf $F_{5GHz}$&\bf SI& \bf$F_{12 \mu{\rm m}}$ &
\bf$F_{25 \mu{\rm m}}$&\bf$F_{60\mu{\rm m}}$&\bf$F_{100 \mu{\rm m}}$& \bf z&\bf $F_{5GHz}^{z=0.1}$\\
(1950 coord)&(1950 coord)& [Jy]& & [Jy]& [Jy]& [Jy]& [Jy]& & [Jy]\\
\hline
&&&&&&&&\\
1Jy 0851+20&IRAS F08519+2017&2.62&0.11&0.2838&0.4335&0.8911&1.1580& 0.306&36.8\\
1Jy 1404+28&IRAS F14047+2841&2.95&0.80&0.1850&0.3994&0.7288&0.9376& 0.077&1.7\\
1Jy 1641+39&IRAS F16413+3954&10.81&0.54&0.1068&0.2918&0.5999&1.3840& 0.594&756\\
1Jy 1308+32&IRAS F13080+3237&1.53&0.20&0.1744&0.2710&0.4198&0.5760& 0.996&525\\
1Jy 2200+42&IRAS F22006+4202&4.77&-0.13&0.1393&0.2495&0.4029&1.9690& 0.070&2.2\\
1Jy 1732+38&IRAS F17326+3859&1.13&0.85&0.0964&0.1688&0.3280&0.6699& 0.970&246\\
1Jy 1803+78&IRAS F18036+7827&2.62&0.25&0.0931&0.1826&0.3083&0.7386& 0.684&301\\
1Jy 0735+17&IRAS F07352+1749&1.99&0.05&0.1868&0.1907&0.2519&0.5198& 0.424&66.3\\
1Jy 0716+71&IRAS F07162+7126&1.12&0.22&0.1121&0.1260&0.2374&0.7825& 0.300&15.7\\
1Jy 0235+16&IRAS F02358+1623&2.85&1.03&0.1105&0.1068&0.2344&0.9074& 0.851&403\\
1Jy 1418+54&IRAS F14180+5437&1.09&0.38&0.0659&0.0857&0.2121&0.5459& 0.151&2.7
\end{tabular}
}
\end{footnotesize}
\vspace{0.5cm}
\caption{\label{IR_table}Strongest IR sources in blazar catalog. $F_{5GHz}$
  stands for the radio flux at $5$ GHz, while the mean IR flux at $12, 25, 60$
  and   $100$ $\mu$m is labeled $F_{12\mu m} \dots F_{100\mu m}$. SI is the
  spectral 
  index $\alpha$ determined between $2.7$ GHz and $5$ GHz. $z$ is the
  cosmological redshift. The distance
  corrected  radio flux flux is given by $F_{5GHz}^{z=0.1}$. 
  The IR data has been taken in the 1980s.
}
\end{table}
\end{center}
\begin{center}
\begin{table}[ht]
\begin{footnotesize}
\centerline{
\begin{tabular}{c|c|c|c|c|c|c|c|c}
 \bf X-ray source&\bf Radio source&\bf $F_X$&\bf HR I&\bf HR II&\bf  $F_{5GHz}$&\bf SI&\bf z&
 \bf $F_{5GHz}^{z=0.1}$\\
(2000 coord)&(1950 coord)&[cnt/s]&&&[Jy]& &\\
\hline
&&&&&&&\\
1RXS J172320.5+341756 &  S4 1721+343 & 1.0160 & 0.20  & 0.07 & 0.934 & -0.43 & 0.206 & 5.2\\
1RXS J225358.0+160855 & 1Jy 2251+158 & 0.6094 & 0.96  & 0.26 & 17.42 & 0.64 & 0.859 & 3100\\
1RXS J084125.1+705342 & 1Jy 0836+710 & 0.5308 & 0.62  & 0.19 &  2.59 & -0.32 & 2.16 & 15300\\
1RXS J115324.4+493108 & 1Jy 1150+497 & 0.3699 & 0.26  & 0.07 &  1.12 & -0.48 & 0.334 & 22.1\\
1RXS J164258.9+394822 & 1Jy 1641+399 & 0.3253 & -0.18 & -0.05 & 10.81 &  0.54 & 0.594 & 756\\
1RXS J192748.0+735757 & 1Jy 1928+738 & 0.2772 & 0.86  & 0.25 &  3.34 & -0.01 & 0.36  & 73.1\\
1RXS J220315.6+314535 & 1Jy 2201+315 & 0.2424 & 0.94  & 0.28 &  2.32 & 0.24 & 0.298 & 29.9\\
1RXS J092702.8+390221 & 1Jy 0923+392 & 0.2376 & -0.16 & 0.38 &  8.73 & 1.03 & 0.698 & 756 \\
1RXS J072153.2+712031 & 1Jy 0716+714 & 0.1925 & 0.12  & 0.19 &  1.12 & 0.22 & 0.30 &14.7\\
1RXS J163813.1+572028 & 1Jy 1637+574 & 0.1158 & 0.03  & 0.24 &  1.42 & 0.35 & 0.750 &202\\
1RXS J074541.2+314249 &  S2 0742+31  & 0.1147 & 0.81  & 0.57 &  0.96 & -0.23 & 0.462 &43.6\\
1RXS J220244.4+421626 & 1Jy 2200+420 & 0.1144 & 0.98  & 0.44 &  4.77 & -0.13 & 0.070 & 2.2\\
1RXS J083454.3+553417 & 1Jy 0831+557 & 0.1027 & 0.58  & 0.03 &  5.65 & -0.46 & 0.242 & 39.7\\
1RXS J135703.6+191915 & 1Jy 1354+195 & 0.1005 & 0.42  & 0.69 &  1.56 & -0.07 & 0.720 & 239 \\
\end{tabular}
}
\end{footnotesize}
\vspace{0.5cm}
\caption{ROSAT sources identified with strong radio blazars.
The mean X-ray flux as measured by ROSAT in 1990/1991 (6 month of observation
time) is given by $F_X$. The
hardness ratios HR I and HR II display the ratio between low and 
high energy X-rays: HR $= (n_{HE}-n_{LE})/(n_{HE}+n_{LE})$. 
For HR I the low energy range is $0.1-0.4$ keV and the 
high energy range is $0.5-2.0$ keV.
For HR II the intervals are $0.5-0.9$ keV and $0.9-2.0$ keV.
SI is the spectral index $\alpha$ determined between $2.7$ GHz and $5$ GHz. $F_{5GHz}$
  stands for the radio flux at $5$ GHz.
SI is the spectral index
$\alpha$ determined between $2.7$ GHz and $5$ GHz
 with $F_v \propto v^\alpha$. The distance
  corrected  radio flux flux is given by $F_{5GHz}^{z=0.1}$.}
\label{rosat_table}
\end{table}
\end{center}
\begin{center}
\begin{table}[htb]
\begin{footnotesize}
\centerline{
\begin{tabular}{c|c|c|c|c|c|c}
\bf HEAO-A source & \bf radio source & \bf  $F_X$ & \bf $F_{5 Ghz}$ & \bf SI& 
\bf z &\bf $F_{5GHz}^{z=0.1}$ \\
(1950 coord)&(1950 coord)& [cnt/s] & [Jy] & & & [Jy]\\
\hline
&&&&&\\
1H0717+714 & 1Jy 0716+714 & 0.0030 & 1.12 & 0.22  & 0.30& 14.7\\
1H1154+294 & S3 1156+29 &   0.0029 &  0.89 & -0.44 &  0.73  & 167\\
1H1922+746 & 1Jy 1928+73 &  0.0027 &   3.34 &  -0.01 &  0.36 & 73.1
\end{tabular}
}
\end{footnotesize}
\vspace{0.5cm}
\caption{HEAO-A sources identified with strong radio blazars: 
The mean flux measured by HEAO-A is given by $F_X$, and  $F_{5GHz}$
  stands for the radio flux at $5$ GHz. All data have been taken within an 6
  month interval in 1977-1978, though the observation times of individual
  sources  are shorter.
SI is the spectral index
$\alpha$ determined between $2.7$ GHz and $5$ GHz
 with $F_v \propto v^\alpha$. The cosmological redshift is given by $z$ and
  the distance
  corrected  radio flux flux is given by $F_{5GHz}^{z=0.1}$. }
\label{heao-a_table}
\end{table}
\end{center}
\begin{table}[htb]
\begin{center}
\begin{footnotesize}
\centerline{
\begin{tabular}{c|c|c|c|c|c|c}
\bf GeV source& \bf radio source&\bf F$_{5 \rm GHz}$&\bf SI
&\bf F$^{\rm max}_{\rm GeV}*10^{-8}$&\bf F$^{\rm mean}_{\rm GeV}*10^{-8}$&\bf z\\
(2000 coord)&(1950 coord)&[Jy]& &$[s^{-1} cm^{-2}]$&$[ s^{-1} cm^{-2}]$\\
\hline
\rm
&&&&&\\
3EG J0450+1105&  1Jy 0446+112& 1.23& 0.56& 109.5& 14.9&1.207  \\
3EG J1635+3813& 1Jy 1633+382& 4.02& 0.73& 107.5& 58.4&1.814  \\
3EG J0530+1323&  1Jy 0528+134& 3.97& 0.47 &93.5 &93.5&2.060  \\
3EG J0237+1635&  1Jy 0235+164& 2.85& 1.03 &65.1& 25.9& 0.940  \\
3EG J2254+1601& 1Jy 2251+158& 17.42& 0.64& 53.7& 53.7&0.859  \\
3EG J1200+2847&S3 1156+29&0.89&-0.44&50.9&7.5&0.73  \\
3EG J2202+4217& 1Jy 2200+420& 4.77&-0.13& 39.9& 11.1&0.069  \\
3EG J1608+1055 &1Jy 1606+106& 1.49& 0.42 &34.9& 25.0&1.226  \\
3EG  J1614+3424& 1Jy 1611+343& 2.67& 0.10& 26.5& 26.5&1.401  \\
3EG  J0829+2413&S3 0827+24&0.94&0.05&24.9&24.9&2.05  \\
3EG  J0204+1458&  1Jy 0202+149& 2.47&-0.43& 23.6& 8.7 & 0.405  \\
3EG  J2232+1147& 1Jy 2230+114& 3.61&-0.50& 19.2& 19.2&1.037  \\
3EG  J1738+5203& 1Jy 1739+522& 1.98& 0.68& 18.2& 18.2&1.375  \\
3EG  J0721+7120& 1Jy 0716+714& 1.12& 0.22 &17.8& 17.8&0.300  \\
3EG  J0737+1721& 1Jy 0735+178& 1.99& 0.05& 16.4& 16.4&0.424  \\
3EG  J0958+6533& 1Jy 0954+658& 1.46& 0.35& 15.4& 6.0&0.368  \\
3EG  J2358+4604& 1Jy 2351+456& 1.42&-0.05& 14.3& 14.3&1.992  \\
3EG  J0239+2815&1Jy 0234+285&1.44&-0.24&13.8&13.8 &1.21  \\
3EG  J0917+4427&S4 0917+449&0.80&0.66&13.8&13.8&2.18  \\
3EG  J0853+1941& 1Jy 0851+202& 2.62& 0.11&10.6& 10.6&0.306   \\
3EG J0845+7049& 1Jy 0836+710& 2.59& -0.32& 10.2& 10.2&2.172  \\
3EG  J0952+5501& 1Jy 0954+556& 2.28&-0.19& 9.1& 9.1&0.901  
\end{tabular}
}
 \end{footnotesize}
 \end{center}
 \vspace{0.5cm}
 \caption{3EG sources identified with strong radio blazars: $F_{5GHz}$ is the
   radio flux at $5$ GHz,
 SI is the spectral index
 $\alpha$ determined between $2.7$ GHz and $5$ GHz. The maximum and the mean
 flux as measured by EGRET in 1991-1995 are listed under F$^{\rm max}_{\rm GeV}$ resp.\
 F$^{\rm mean}_{\rm GeV }$. $z$ is the
   cosmological redshift. The EGRET catalog uses the reference system of the yea
   2000.} 
 \label{egret_table}
 \end{table}
 \begin{table}[pht]
 \begin{center}
 \begin{footnotesize}
\centerline{
\begin{tabular}{c|c|c|c|c}
\bf EGRET source& \bf $F_{GeV}^{max} $ & \bf $F_{GeV}^{mean}$& $|b|$ & 3EG Id.\ flag\\
&$[ 10^{-8} s^{-1} cm^{-2} ]$&$[ 10^{-8} s^{-1} cm^{-2} ]$&$[^\circ]$&\\
\hline
&&&&\\
3EG J2243+1509 & 73.1 & 9.9 & 37.49 &  \\
3EG J2255+1943 & 62.2 & 5.8 & 35.43 & a \\
3EG J1835+5918 & 60.6 & 60.6 & 25.07 &  \\
3EG J1212+2304 & 50.8 & 3.3 & 80.32 &  \\
3EG J1850+5903 & 46.7 & 12.6 & 23.18 &  \\
3EG J0439+1555 & 42.9 & 4.8 & 19.98 &  \\
3EG J0010+7309 & 42.3 & 42.3 & 10.54 &  \\
3EG J1822+1641 & 40.6 & 7.1 & 13.84 &  \\
3EG J2314+4426 & 40.4 & 13.9 & 15.10 &  \\
3EG J2352+3752 & 37.5 & 6.1 & 23.54 & a \\
3EG J1825+2854 & 34.3 & 6.5 & 18.03 &  \\
3EG J0407+1710 & 32.1 & 7.3 & 25.06 &  \\
3EG J1824+3441 & 28.7 & 8.1 & 20.14 &  \\
3EG J1308+8744 & 23.9 & 7.6 & 29.38 &  \\
3EG J1733+6017 & 22.9 & 8.7 & 32.94 &  \\
3EG J1227+4302 & 21.7 & 4.6 & 73.33 &  \\
3EG J1347+2932 & 21.0 & 9.6 & 77.50 &  \\
3EG J0910+6556 & 18.3 & 5.9 & 38.56 &  \\
3EG J1323+2200 & 18.1 & 5.2 & 81.15 & a \\
3EG J0215+1123 & 18.0 & 9.3 & 46.37 &  \\
3EG J0329+2149 & 17.2 & 7.4 & 27.88 &  \\
3EG J0245+1758 & 16.9 & 8.8 & 37.11 &  \\
3EG J0426+1333 & 14.0 & 14.0 & 23.82 &  \\
3EG J0917+4427 & 13.8 & 13.8 & 44.19 & a \\
3EG J2248+1745 & 12.9 & 12.9 & 36.17 &  \\
\end{tabular}
}
\end{footnotesize}
\end{center}
\vspace{0.5cm}
\caption{
Unidentified and weakly identified extragalactic EGRET sources, 
possibly blazars. SI is the spectral index
$\alpha$ determined between $2.7$ GHz and $5$ GHz. $F_{5GHz}$ is the
  radio flux at $5$ GHz,
SI is the spectral index
$\alpha$ determined between $2.7$ GHz and $5$ GHz. The maximum and the mean
flux as measured by EGRET in 1991-1995 are listed under F$^{\rm max}_{\rm GeV}$ resp.\
F$^{\rm mean}_{\rm GeV }$.
\label{3eg_unid_table}
}
\end{table}
\begin{center}
\begin{table}[htb]
\begin{footnotesize}
\centerline{
\begin{tabular}{c|c|c|c|c}
\bf X-ray source& \bf other name&  \bf $F_{TeV}$& \bf 5 GHz flux& \bf z\\
& &[crab flux]&[Jy]&\\
\hline
&&&\\
1ES 1101+38.4&Mrk 421&0.04-7.4&0.72&0.0300\\
1ES 1652+39.8&Mrk 501&0.33-6&1.38&0.0336\\
H 1426+428&-&0.03-0.08&0.04& 0.1290\\
1ES 1959+650&-&0.05-2.20&0.25&0.0470\\
1ES 2344+514&-&0.03&0.23&0.0440
\end{tabular}
}
\end{footnotesize}
\vspace{0.5cm}
\caption{TeV blazars with $\delta>10^\circ$,
The TeV fluxes are listed in units of  the
flux of the Crab nebula, $dJ_\gamma/dE = (2.79\pm 0.02 \pm 0.5) \cdot 10^{-7}
(E/TeV)^{-2.59\pm0.03\pm0.05}$. $F_{5GHz}$
  stands for the radio flux at $5$ GHz, while the cosmological redshift is
  given by $z$.} 
\label{TeV_table}
\end{table}
\end{center}
\begin{center}
\begin{table}[!h]
\begin{small}
\centerline{
\small
\begin{tabular}{c|c|c|c|c|c|c|c|c|c}
\bf Coordinate & \bf other  &\bf Sample &\bf Mag&\bf z&\bf $F_{5GHz}$&\bf $\theta$
&\bf $\nu_m$&\bf $\log{P_{5GHz}}$&\bf Size\\
\bf name&\bf name&&&&&&&&\\
(1950 coord)&&&&& [Jy]  & [arcsec]& [MHz]&&[kpc]\\
\hline 
0248+430 & n. a.        & G &  15.5 & 1.316 & 1.24 & 0.06 & 5000 & 27.9 & 0.474 \\
0134+329 & 3C 48        & C &  16.1 & 0.367 & 5.3 & 0.50 & 80 & 27.3 & 2.255  \\
0738+313 & n. a.        & G &  16.1 & 0.630 & 3.62 & 0.010 & 5300 & 27.6 & 0.061 \\
1458+718 & 3C 309.1     & C &  16.8 & 0.905 & 3.5 & 2.11 & 40 & 28.0 & 14.831  \\
1345+125 & 4C 12.50     & G &  17.0 & 0.122 & 3.05 & 0.080 & 400 & 26.0 & 0.160 \\
1328+307 & 3C 286       & C &  17.2 & 0.849 & 7.4 & 3.2 & 80 & 28.2 & 21.966  \\
0740+380 & 3C 186       & C &  17.6 & 1.063 & 0.3 & 2.2 & 40 & 27.2 & 16.330  \\
1328+254 & 3C 287       & C &  17.7 & 1.055 & 3.2 & 0.048 & 50 & 28.1 & 0.355 \\
0538+498 & 3C 147       & C &  17.8 & 0.545 & 8.2 & 0.55 & 150 & 27.9 & 3.101 \\
1442+101 & OQ 172       & CG & 17.8 & 3.535 & 1.20 & 0.02 & 900 & 29.3 & 0.185 \\
0221+276 & 3C 67        & C &  18.0 & 0.309 & 0.9 & 2.5 & 50 & 26.3 & 10.100  \\
2352+495 & n. a.        & G &  18.4 & 0.237 & 1.49 & 0.050 & 700 & 26.3 & 0.168  \\
2249+185 & 3C 454       & C &  18.5 & 1.758 & 0.8 & 0.66 & 40 & 28.1 & 5.562  \\
0518+165 & 3C 138       & C &  18.8 & 0.760 & 4.1 & 0.60 & 100 & 27.8 & 3.942 \\
0345+337 & 3C 93.1      & C &  19.0 & 0.244 & 0.8 & 0.25 & 60 & 26.0 & 0.858  \\
1153+317 & 4C 31.38     & C &  19.0 & 1.557 & 1.0 & 0.9 & 100 & 28.0 & 7.397  \\
1203+645 & 3C 268.3     & C &  19.0 & 0.371 & 1.1 & 1.36 & 80 & 26.6 & 6.175  \\
1323+321 & 4C 32.44     & CG & 19.0 & 0.369 & 2.39 & 0.06 & 500 & 26.9 & 0.272  \\
1443+77 & 3C 303.1      & C &  19.0 & 0.267 & 0.5 & 1.7 & 100 & 26.0 & 6.219  \\
1819+39 & 4C 39.56      & C &  19.0 & 0.4 & 1.0 & 0.44 & 100 & 26.6 & 2.091  \\
1358+624 & 4C 62.22     & CG & 19.9 & 0.429 & 1.80 & 0.07 & 500 & 26.9 & 0.347 \\
0127+233 & 3C 43        & C &  20.0 & 1.459 & 1.1 & 2.60 & 20 & 28.0 & 21.054  \\
0428+205 & OF 247       & CG & 20.0 & 0.219 & 2.38 & 0.250 & 1100 & 26.4 & 0.793 \\
1117+146 & 4C 14.41     & G &  20.0 & 0.362 & 1.00 & 0.08 & 500 & 26.5 & 0.358  \\
1829+29 & 4C 29.56      & C &  20.0 & 0.842 & 1.1 & 3.1 & 100 & 27.3 & 21.212  \\
\end{tabular}
}
\end{small}
\vspace{0.5cm}
\caption{GPS and CSS sources fulfilling our selection rules. In Sample, G
  means GPS and  C means CSS. Mag is the optical magnitude. The cosmological
redshift is given by $z$, while $F_{5GHz}$
  stands for the radio flux at $5$ GHz.
$\theta$ is the  angular size of the source. $\nu_m$ is the  
turnover frequency and $P_{5GHz}$ stands for $\log$ of power at 5 GHz (assuming isotropic
emission). The linear size of the sources is listed in kpc. The data were
  compiled from previous catalogs in 1998.
 }
\label{table_CSSGPS}
\end{table}
\end{center}

\begin{center}
\begin{table}[!h]
\begin{footnotesize}
\centerline{
\begin{tabular}{c|c|c|c|c|c}
\bf Source & \bf $F_{178MHz}$ &\bf SI& \bf $\delta$ & \bf RA & \bf z\\
&[Jy]&&[ $^\circ$] & [ $^\circ$]&\\
\hline
&&&&&\\
3C 274.0 & 1050. & -0.76 & 12.6667 & 12.4667 & 0.004 \\
3C 84.0 & 61.3 & -0.78 & 41.3167 & 3.26667 & 0.017 \\
3C 433.0 & 56.2 & -0.75 & 24.85 & 21.35 & 0.101 \\
3C 310.0 & 55.1 & -0.92 & 26.2 & 15.0333 & 0.054 \\
3C 338.0 & 46.9 & -1.19 & 39.65 & 16.4333 & 0.029 \\
3C 465.0 & 37.8 & -0.75 & 26.75 & 23.5833 & 0.029 \\
3C 83.1 & 26.0 & -0.64 & 41.6667 & 3.23333 & 0.025 \\
3C 264.0 & 26.0 & -0.75 & 19.8833 & 11.7 & 0.020 \\
3C 66.0 & 24.6 & -0.62 & 42.75 & 2.33333 & 0.021 \\
3C 386.0 & 23.9 & -0.59 & 17.15 & 18.6 & 0.017 \\
3C 272.1 & 19.4 & -0.60 & 13.15 & 12.3667 & 0.003 \\
3C 288.0 & 18.9 & -0.85 & 39.1 & 13.6 & 0.246 \\
3C 315.0 & 17.8 & -0.72 & 26.3 & 15.1833 & 0.108 \\
3C 31.0 & 16.8 & -0.57 & 32.1333 & 1.06667 & 0.016 \\
3C 28.0 & 16.3 & -1.06 & 26.1333 & 0.883333 & 0.195 \\
3C 442.0 & 16.1 & -0.96 & 13.5833 & 22.2 & 0.026 \\
3C 305.0 & 15.7 & -0.85 & 63.4667 & 14.8 & 0.041 \\
3C 231.0 & 14.6 & -0.28 & 69.9167 & 9.85 & 0.000 \\
3C 296.0 & 13.0 & -0.67 & 11.0333 & 14.2333 & 0.023 \\
3C 293.0 & 12.7 & -0.45 & 31.6833 & 13.8333 & 0.045 \\
3C 76.1 & 12.2 & -0.77 & 16.2333 & 3 & 0.032 \\
3C 449.0 & 11.5 & -0.58 & 39.1 & 22.4833 & 0.017 \\
3C 346.0 & 10.9 & -0.52 & 17.35 & 16.6833 & 0.161 \\
3C 314.1 & 10.6 & -0.95 & 70.95 & 15.1667 & 0.119 \\
\end{tabular}
}
\end{footnotesize}
\vspace{0.5cm}
\caption{FR-I radio galaxies: $F_{178MHz}$ is the mean radio flux at 178 MHz,
SI is the spectral index $\alpha$ determined between 178 MHz and 750 MHz,
 RA and $\delta$ stand for
right ascension and declination, respectively in the reference system of 1950.
All data were collected before 1985.
Note, in the original 3CRR catalog, 
the spectral index is defined with an opposite sign. The cosmological redshift
 is given by  $z$. 
}
\label{FRI}
\end{table}
\end{center}
\begin{center}
\begin{table}[hp]
\begin{footnotesize}
\centerline{
\begin{tabular}{c|c|c|c|c|c}
\bf Source & \bf $F_{178MHz}$ &\bf SI& \bf $\delta$ & \bf RA & \bf z\\
&[Jy]&&[ $^\circ$]  &[ $^\circ$]&\\
\hline
&&&&&\\
3C 123.0 & 189.0 & -0.70 & 29.5667 & 4.55 & 0.218 \\
3C 295.0 & 83.5 & -0.63 & 52.4333 & 14.15 & 0.461 \\
3C 196.0 & 68.2 & -0.79 & 48.3667 & 8.15 & 0.871 \\
3C 452.0 & 54.4 & -0.78 & 39.4167 & 22.72 & 0.081 \\
3C 33.0 & 54.4 & -0.76 & 13.0667 & 1.1 & 0.059 \\
3C 390.3 & 47.5 & -0.75 & 79.7167 & 18.75 & 0.056 \\
3C 98.0 & 47.2 & -0.78 & 10.2833 & 3.93 & 0.030 \\
3C 438.0 & 44.7 & -0.88 & 37.7667 & 21.88 & 0.290 \\
3C 20.0 & 42.9 & -0.67 & 51.7833 & 0.67 & 0.174 \\
3C 219.0 & 41.2 & -0.81 & 45.85 & 9.28 & 0.174 \\
3C 234.0 & 31.4 & -0.86 & 29.0167 & 9.97 & 0.184 \\
3C 61.1 & 31.2 & -0.77 & 86.0833 & 2.17 & 0.186 \\
3C 79.0 & 30.5 & -0.92 & 16.9 & 3.12 & 0.255 \\
3C 330.0 & 27.8 & -0.71 & 66.0667 & 16.15 & 0.550 \\
3C 427.1 & 26.6 & -0.97 & 76.35 & 21.07 & 0.572 \\
3C 47.0 & 26.4 & -0.98 & 20.7 & 1.55 & 0.425 \\
3C 388.0 & 24.6 & -0.70 & 45.5 & 18.7 & 0.090 \\
3C 280.0 & 23.7 & -0.81 & 47.6 & 12.9 & 0.997 \\
3C 228.0 & 21.8 & -1.00 & 14.5667 & 9.78 & 0.552 \\
3C 109.0 & 21.6 & -0.85 & 11.0667 & 4.17 & 0.305 \\
3C 55.0 & 21.5 & -1.04 & 28.6167 & 1.9 & 0.734 \\
3C 268.1 & 21.4 & -0.59 & 73.2833 & 11.95 & 0.973 \\
3C 225.0B & 21.3 & -0.94 & 13.9833 & 9.65 & 0.582 \\
3C 192.0 & 21.1 & -0.79 & 24.3 & 8.03 & 0.059 \\
3C 401.0 & 20.9 & -0.71 & 60.5667 & 19.65 & 0.201 \\
\end{tabular}
}
\end{footnotesize}
\vspace{0.5cm}
\caption{FR-II radio galaxies: $F_{178MHz}$ is the mean radio flux at 178 MHz,
SI is the spectral index $\alpha$ determined between 178 MHz and 750 MHz
 RA and $\delta$ stand for
right ascension and declination, respectvely in the reference system of 1950.
All data were collected before 1985.
Note, in the original 3CRR catalog, 
the spectral index is defined with an opposite sign. The cosmological redshift
 is given by  $z$. 
}
\label{FRII}
\end{table}
\end{center}
\begin{center}
\begin{table}[!h]
\begin{footnotesize}
\centerline{
\begin{tabular}{c|c|c|c}
quasar name& $F_{60\mu m}$& z & $\log \nu f_{14.5}$\\
  & [mJy] &&\\
\hline
&&&\\
PG 0050+124 & 2293 & 0.061 & -13.30  \\
PG 1351+640 & 757  & 0.087 & -13.86  \\
PG 1440+356 & 652  & 0.077 & -13.70  \\
PG 1613+658 & 635  & 0.129 & -13.97  \\
PG 1119+120 & 546  & 0.049 & -13.82  \\
PG 1501+106 & 486  & 0.036 & -13.54  \\
PG 1700+518 & 480  & 0.292 & -13.90  \\
PG 1351+236 & 364  & 0.055 & -13.75  \\
PG 1543+489 & 348  & 0.400 & -14.52  \\
PG 2214+139 & 337  & 0.067 & -13.70  \\
PG 1634+706 & 318  & 1.334 & -14.00  \\
PG 1402+261 & 229  & 0.164 & -14.10  \\
PG 1248+401 & 224  & 1.030 & -14.59  \\
PG 0947+396 & 201  & 0.206 & -14.57  \\
PG 1148+549 & 196  & 0.969 & -14.30  \\
PG 1114+445 & 191  & 0.144 & -14.14  \\
PG 0804+761 & 191  & 0.100 & -13.70  \\
PG 0838+770 & 174  & 0.131 & -14.25  \\
PG 0906+484 & 172  & 0.118 & -14.30  \\
PG 1229+204 & 163  & 0.064 & -13.74  \\
PG 0844+349 & 163  & 0.064 & -13.82  \\
PG 1411+442 & 162  & 0.089 & -13.87  \\
PG 1448+273 & 117  & 0.065 & -13.81  \\
PG 1444+407 & 117  & 0.267 & -14.28  \\
PG 1415+451 & 112  & 0.114 & -14.05  \\
PG 0052+251 & 93   & 0.155 & -14.09  \\
\end{tabular}
}
\end{footnotesize}
\vspace{0.5cm}
\caption{Radio-weak quasars from the BQS sorted according to IRAS mean
flux at $\lambda=60\mu$m, F$_{60\mu m}$. The cosmological redshift
 is given by  $z$. The logarithm of the energy flux density at $10^{14.5}$ is given by $\log
 \nu f_{14.5}$. For the BQS, data were taken until 1983, IRAS data have
 been taken in the 1980s.  
}
\label{table_bqs}
\end{table}
\end{center}

\begin{center}
\begin{table}[!h]
\centerline{
\begin{tabular}{c|c|c}
\bf Radio source&\bf  other name&\bf Samples (Position)\\
\hline
&&\\
1Jy 0316+41&NGC 1275&IR blazars (1), keV blazars (ROSAT (1) and HEAO-A (1))\\
1Jy 1652+398&Mrk 501& keV blazars (ROSAT (2) and HEAO-A (2))\\
1Jy 1222+13&M 84& IR blazars (6)\\
1Jy 0055+30&NGC 315 & IR blazars (11)\\
1Jy 1807+698& &keV blazars (HEAO-A (31))\\
\end{tabular}
}
\vspace{0.5cm}
\caption{
\label{table_nearby}
Nearby intrinsically weak sources. Due to their  proximity
they nevertheless may contribute
significantly to the neutrino flux. We propose to analyze them as individual
sources.}
\end{table}
\end{center}

\end{document}